\documentclass[12pt]{article}

\newcommand{\al}{\alpha}

\newcommand{\bal}{{\bar \alpha}} 
\newcommand{\ice}[1]{\relax}
\newcommand{\be}{\begin{equation}}
\newcommand{\ee}{\end{equation}}
\newcommand{\bea}{\begin{eqnarray}}
\newcommand{\eea}{\end{eqnarray}}
\newcommand{\nn}{\nonumber}
\newcommand{\MSsch}{{\overline{\rm MS}}}
\begin{document}
\thispagestyle{empty}
\begin{flushright}
MZ-TH/00-51\\
November 2000\\
\end{flushright}
\vspace{0.5cm}
\begin{center}
{\Large\bf Running electromagnetic coupling constant: \\
low energy normalization and the value at $\bf M_Z$}\\[1truecm]
{\large {\bf A.A.~Pivovarov}}\\[.1cm]
Institut f\"ur Physik, Johannes-Gutenberg-Universit\"at,\\[-.1truecm]
Staudinger Weg 7, D-55099 Mainz, Germany\\
and\\
Institute for Nuclear Research of the\\[-.1truecm]
Russian Academy of Sciences, Moscow 117312, Russia
\vspace{0.6truecm}
\end{center}

\begin{abstract}
\noindent A numerical value for the 
running electromagnetic coupling constant in the $\MSsch$ scheme
is calculated 
at the low energy normalization scale equal to the $\tau$-lepton 
mass $M_\tau$.
This low energy boundary value is used
for running the electromagnetic coupling constant
to larger scales where high precision 
experimental measurements can be performed. Particular scales
of interest are the $b$-quark mass for studying 
the $\Upsilon$-resonance physics 
and the $Z$-boson mass $M_Z$ for high precision 
tests of the standard model and for the Higgs mass
determination from radiative corrections. 
A numerical value for the 
running electromagnetic coupling constant
at $M_Z$ in the on-shell renormalization scheme is also given.
\end{abstract}
\newpage

\section{Introduction.}
Dimensional regularization (DR) and minimal subtraction (MS) are
convenient and widely used
technical tools for perturbative calculations in particle
phenomenology \cite{DR,MS1}. In low orders of perturbation theory (PT) 
dimensional regularization does not give any decisive computational advantage.
However, the high order PT many-loop calculations 
are rather involved and, in practice, only 
dimensional regularization supplemented by the recurrence relations 
based on the integration-by-part technique \cite{intbypart}
allowed one to obtain new analytical results, e.g. \cite{chetPhRep}.
Minimal subtraction, being a simple method of renormalizing  
the dimensionally regularized PT diagrams, 
is now becoming a dominant way of theory parameterization 
in a form of the $\MSsch$ scheme \cite{MS2}.  
The renormalization in the $\MSsch$ scheme is mass
independent that allows an efficient computation of renormalization
group functions describing scaling of $\MSsch$ parameters.
However, the mass independence of the renormalization procedure is
physically inconvenient because decoupling of heavy particles 
is not automatic \cite{decoupl}. 
The physical property of decoupling is restored within
an effective theory approach 
with the explicit separation of different mass scales such that 
the parameters of neighboring effective theories 
(couplings, masses, ...) should be sewed (matched) near 
the point where a new scale appears.
This machinery, worked out up to three-loop order in PT,
allows one to compare theoretical results in the $\MSsch$ scheme for a
variety of scales with a uniform control
over the precision of PT calculation. In particular, 
this technique allows one to
compare theoretical quantities extracted from 
the low energy data with results of the $Z$-boson peak analyses
within the standard model (SM) of particle interactions.
The high precision tests of SM at the $Z$-boson peak,
completed at one-loop level, have shown a good agreement with 
theoretical results obtained from the low energy data.
For the search of new physics and further
tests of SM at the next level of precision the computations 
for many observables at the $Z$-boson peak 
should be done with two-loop accuracy which presently is an actual 
calculational task. 
Because of the computational advantage of dimensional regularization
in many-loop calculations, the high order PT results 
for theoretical amplitudes at the $Z$-boson peak tend
to be obtained in terms of the $\MSsch$ scheme
parameters which are natural quantities for 
the minimally-subtracted 
dimensionally-regularized diagrams. 
It was found that the use of the running electromagnetic (EM) 
coupling normalized at $M_Z$ 
in the $\MSsch$ scheme makes PT expansions 
near the $Z$-boson peak reliable and corrections small.
However, contrary to the fine structure constant $\al$,
the running EM coupling in the $\MSsch$ scheme
has no immediate physical meaning and its
numerical value is not well known. 
At the same time, 
QED, being an old part of the SM,
is well tested at low energies where 
the fine structure constant $\al$ is a natural 
interaction parameter defined 
in a physical manner by subtraction on the photon mass shell.
The fine structure constant is very well known numerically
that would make it a natural reference parameter for high precision tests 
of SM. However, because of a huge numerical difference
between the values of the photon and $Z$-boson masses
the use of the fine structure constant 
for calculations at the $Z$-boson peak
generates large corrections in PT.
For applications to high precision tests of the standard model
with observables near the $Z$-boson peak \cite{SMrev},
one should transform $\al$ into a proper high energy 
parameter, i.e. 
into the electromagnetic coupling constant at a scale 
of the $Z$-boson mass $M_Z$ (see e.g. \cite{CERNwork,Hollik}).
Then the large PT corrections are hidden (renormalized)
into a numerical value of this new parameter
which is more natural for describing 
the $Z$-boson peak observables than $\al$.
Therefore, a numerical value of 
the running EM coupling constant at $M_Z$
is a new important number which has been chosen
as a standard reference parameter \cite{PDG}. 
A difference of the numerical value for this parameter from 
$\alpha^{-1}=137.036\ldots$
should be theoretically calculated.
The change is accounted for through the renormalization group
(RG) technique \cite{RG,GellMannLow,RGeqBS}.
Because the fine structure constant is defined
at vanishing momentum it is an infrared sensitive quantity
and a contribution of strong interactions 
into its RG evolution cannot be computed perturbatively: 
the infrared 
region is a domain of strong coupling that requires 
a nonperturbative (nonPT) treatment.
The contribution of the infrared (IR) region
is usually taken into account within a semi-phenomenological
approximation through a dispersion relation
with direct integration of low energy data.
There has been a renewal of interest in 
a precise determination of the hadronic
contribution into 
the electromagnetic coupling constant at $M_Z$ 
during the last years in connection with
the constraints on the Higgs boson mass from radiative corrections in SM
\cite{Higgs}. 
Some recent references giving
a state-of-the-art analysis of this contribution
are
\cite{Davier1,Schilcher,Kuhn,Davier}. 
A quasi-analytical approach was used in 
ref.~\cite{Krasnikov} where some references to earlier papers can be found
(see also \cite{ITEP,burkhardt}).
An extremely thorough 
data-based analysis is given in ref.~\cite{Jegerlehner}.
However, the virtual 
lack of data for energies higher than 
$15\div 20~{\rm GeV}$ makes it unavoidable to use 
theoretical formulas in the dispersion relation.
Fortunately, the theoretical results necessary for electromagnetic
current correlators (the photon vacuum polarization function)
are known in high orders of PT 
and are reliable at large energies.
Therefore, the real value of dispersion relations is to find a
boundary condition for the running EM coupling at 
a low energy normalization scale
where data are accurate. If this low energy 
normalization scale is large enough for strong interaction PT
to be applicable then the renormalization
group can be used 
to run the initial value to any larger 
scale with very high precision.  
The running of the electromagnetic coupling constant
can be defined in different ways
depending on the renormalization procedure chosen.
The evolution can be described in both on-shell
and $\MSsch$ schemes: the corresponding 
$\beta$-functions are available with high precision.
The recent calculation of the numerical value 
for the running EM coupling at $M_Z$ with 
evolution in the $\MSsch$ scheme has been presented in ref.~\cite{erler}. 

In the present paper I calculate a low energy boundary condition 
for the running electromagnetic coupling constant
in the $\MSsch$ scheme 
using almost no experimental data
but masses of ground states in the $\rho$- and $\varphi$-meson 
channels. The necessary IR modification of the light quark spectrum
is determined by consistency with OPE.
The theoretical parameters of the calculation 
are the strong coupling constant $\al_s(M_\tau)$,
the strange quark mass $m_s(M_\tau)$ 
and the gluon and quark vacuum condensates. The numerical values for 
these parameters
accumulate a lot of information on low energy data presented in the
standard rate $R(s)$ of $e^+ e^-$ annihilation into hadrons.
Therefore, the present 
calculation compresses low energy data into the numerical values of
several key theoretical parameters that allows one to perform an 
analysis of the IR domain necessary for determination
of the low energy boundary value for the running EM coupling. 
The evolution to larger scales is straightforward and very precise
within perturbation theory.

\section{Basic relations}
The relation between the running EM coupling constant $\bal(\mu)$
in the $\MSsch$ scheme 
and the fine structure constant $\al$
is obtained by considering the photon vacuum polarization function. 
The correlator of the EM current $j_\mu^{em}$
\be
\label{emcorrel}
12\pi^2 i \int \langle T j_\mu^{em}(x)j_\nu^{em}(0) \rangle 
e^{iqx}dx=(q_\mu q_\nu -g_{\mu\nu}q^2)\Pi_{\#}(q^2)
\ee
is defined with a generic scalar function
$\Pi_{\#}(q^2)$.
The particular scalar functions $\Pi(\mu^2,q^2)$ 
and $\Pi_{os}(q^2)$ 
are defined through the correlator
of electromagnetic currents eq.~(\ref{emcorrel})
(and the generic function $\Pi_{\#}(q^2)$)
but with different subtraction procedures to remove infinities.
The first function $\Pi(\mu^2,q^2)$
is renormalized in the $\MSsch$ scheme and the second function 
$\Pi_{os}(q^2)$ is renormalized by subtraction on the
photon mass-shell $q^2=0$ which implies the normalization 
condition $\Pi_{os}(0)=0$. 
Note that for the actual calculation of $\Pi_{os}(q^2)$
one can use dimensional regularization and  
the $\MSsch$ scheme in cases when $\Pi(\mu^2,0)$ exists,
\[
\Pi_{os}(q^2)=\Pi(\mu^2,q^2)-\Pi(\mu^2,0) \, .
\]
The relation between couplings and 
polarization functions in different schemes
reads
\begin{equation}
  \label{match}
\frac{3\pi}{\bal(\mu^2)}+\Pi(\mu^2,q^2)
=\frac{3\pi}{\al}+\Pi_{os}(q^2)\, .
\end{equation}
In the limit $q^2\to 0$ one finds
\begin{equation}
  \label{match0}
\frac{3\pi}{\bal(\mu^2)}+\Pi(\mu^2,0)
=\frac{3\pi}{\al}\, .
\end{equation}
Eq.~(\ref{match}) is related to 
the Coulomb law for charged particles. For the potential of
the EM interaction of two charged leptons
one finds in the $\MSsch$ scheme
\begin{equation}
\label{defCoulMS}
V({\bf q}^2)=-\frac{4\pi\bal(\mu^2)}{{\bf q}^2}
\frac{1}{1+\frac{\bal(\mu^2)}{3\pi}\Pi(\mu^2,{\bf q}^2)}\, .
\end{equation}
This expression is $\mu$ independent because of RG invariance.
Being expressed through the fine structure constant $\al$
the Coulomb potential reads
\begin{equation}
\label{defCoulOS}
V({\bf q}^2)=-\frac{4\pi\al}{{\bf q}^2}
\frac{1}{1+\frac{\al}{3\pi} \Pi_{os}({\bf q}^2)}
\end{equation}
with $\Pi_{os}(0)=0$. The limit of large distances
\be
\label{defCoullim}
4\pi \al=-\lim_{{\bf q}^2\to 0} {{\bf q}^2}V({\bf q}^2)
\ee
gives the fine structure constant.
In the Coulomb law 
eqs.~(\ref{defCoulMS},\ref{defCoulOS})
$q=(0,{\bf q})$ and $q^2=-{\bf q}^2$. This makes $q^2$ in 
eq.~(\ref{match}) essentially Euclidean. We keep notation 
${\bf q}^2$ for a positive number to stress the calculation in 
the Euclidean domain. 
Eq.~(\ref{match0}) is just a 
relation between schemes in which the EM coupling is defined
or finite renormalization.

For massless quarks the limit ${\bf q^2}\to 0$ in
eq.~(\ref{match}) requires care.
The polarization function $\Pi(\mu^2,0)$
cannot be calculated in PT if strong interactions are included
because light quarks are essentially massless.

Besides the $\MSsch$ running coupling constant $\bal(\mu)$,
the on-shell running coupling $\al_{os}({\bf q}^2)$,
which can also be used in the $Z$-boson peak analyses,
is defined through 
\be
\label{OSrun}
\al_{os}({\bf q}^2)=\frac{\al}
{1+\frac{\al}{3\pi} \Pi_{os}({\bf q}^2)}, \quad \al_{os}(0)=\al\, .
\ee
The numerical value for the on-shell running coupling $\al_{os}({\bf q}^2)$
can be found from eq.~(\ref{match}) if $\bal(\mu)$ is known
and $\Pi(\mu^2,{\bf q}^2)$ is calculable for a given ${\bf q}^2$.

In the present paper I calculate the low energy boundary condition 
for the running EM coupling in the $\MSsch$ scheme, i.e. the value  
$\bal(\mu_0)$ at some $\mu_0$. A convenient scale 
is the $\tau$ lepton mass $M_\tau$
which is large enough for strong interaction PT to work,
i.e. $\mu_0=M_\tau$.
The value $\bal(M_\tau)$
can then be run to other scales with the standard RG equation.
The particular values of
interest are $m_b$ for $\Upsilon$ physics and  
$M_Z$ for high precision SM tests and Higgs boson search.
The RG functions in the $\MSsch$ scheme are known
with a very high accuracy that makes the running
very precise numerically.

\section{Low energy normalization: formulas}
One needs a numerical value of the polarization function 
$\Pi(\mu^2,q^2)$ at $q^2=0$ at some low 
normalization point $\mu^2$.
There are lepton and quark contributions to the EM current
(see a note about $W$ bosons below).
Because decoupling is not explicit we count only contributions of
particles which are considered active for a given scale.

\subsection{Leptons}
For a lepton 'l' with the pole mass $M_l$ 
we retain masses that makes $\Pi(\mu^2,0)$
directly computable in low orders of PT
where strong interactions are absent. 
The matching condition reads
\be
\Pi^l(\mu^2,0)
=\ln \frac{\mu^2}{M_l^2}+\frac{\bal(\mu^2)}{\pi}\left(\frac{45}{16}
+\frac{3}{4}\ln\frac{\mu^2}{M_l^2}\right)+{\cal O}(\bal^2)\, .
\label{matlep}
\ee
Note that ${\cal O}(\al^2)$ corrections are also available 
\cite{david} but they are totally negligible numerically for our purposes.
With accuracy of order $\al$ there is no numerical difference between 
the fine structure constant $\al$ 
and the running coupling constant $\bal(\mu^2)$ 
in the RHS of eq.~(\ref{matlep}).
For numerical estimates we substitute $\al$. 
For $\mu=M_l$ the lepton 'l' decouples completely
in the leading order (which can be practical for the $\tau$ lepton).
Because the fine structure constant $\al$ 
is small numerically we do not resum the expression in the RHS of 
eq.~(\ref{matlep}). 
Note that the expression~(\ref{matlep}) 
can basically be used at any $\mu$.
In a sense the matching for leptons 
can be done just at any scale of interest, 
for instance, at $\mu=M_Z$.  
For a lepton we use the pole mass $M_l$
which is a meaningful parameter in finite order PT.
Eq.~(\ref{matlep}) gives the leptonic part of finite renormalization
between the running and fine structure constants in eq.~(\ref{match0}).
In eq.~(\ref{matlep})
we neglect strong interactions (quark contributions) which appear in 
${\cal O}(\al^2)$ order. 
If strong interactions are included then one cannot use PT with such a low
scale as the electron or muon mass and the full IR analysis analogous to
that done for light quarks (see below) is necessary.
Eq.~(\ref{matlep}) solves the lepton part of the normalization condition.

\subsection{Light quarks}
For the hadronic part of the vacuum polarization function 
we first consider a light quark contribution 
which is most complicated.
For the light (massless) quarks the limit ${\bf q^2}\to 0$ in
eq.~(\ref{match}) necessary to relate the running coupling to 
the fine structure constant cannot be performed in PT.
This is, however, an infrared (IR) problem which is unsolved 
in QCD within PT.
The low energy domain is not described in QCD with massless quarks
by PT means 
and PT expressions should be modified for the limit ${\bf q^2}\to 0$
in eq.~(\ref{match}) to exist.
Such a modification must not change an ultraviolet (UV) structure 
of the correlators because RG invariance
should be respected. Therefore, 
it is convenient to perform an IR modification using 
dispersion relations which give contributions of different 
energy ranges separately.
There are three potentially 
IR dangerous quarks $u$, $d$, and $s$. For matching light quarks
we work in $n_f=3$ effective theory, i.e. in QCD with three active light
quarks. 

A note about notation is in order. 
We consider a generic light quark correlator normalized at the parton
level to 1 (as for its asymptotic spectral density).
Then we add necessary factors to account for color and/or charge
structure.
Thus, for $u$ quark, for instance, 
\begin{equation}
  \label{gene}
\Pi^{u}({\bf q}^2)=N_c e_u^2 \Pi^{light}({\bf q}^2)
\end{equation}
where $e_u=2/3$ is a $u$-quark EM charge and $N_c=3$ is a number of colors.
For light quarks the PT part of the correlator is calculable 
for large ${\bf q}^2$ and reads in the $\MSsch$ scheme (e.g. \cite{surgu})
\[
\Pi^{light}(\mu^2,{\bf q}^2)
=\ln \frac{\mu^2}{{\bf q}^2}+\frac{5}{3}
+a_s\left(\ln \frac{\mu^2}{{\bf q}^2}+\frac{55}{12}-4\zeta(3)\right)
\]
\be
\label{Pilight}
+a_s^2\left(\frac{9}{8}\ln^2 \frac{\mu^2}{{\bf q}^2}
+\left(\frac{299}{24}-9\zeta(3)\right)\ln \frac{\mu^2}{{\bf q}^2}
+\frac{34525}{864}-\frac{715}{18}\zeta(3)+\frac{25}{3}\zeta(5)\right)
\ee
where $a_s=\al_s/\pi$, $a_s\equiv a_s^{(3)}(\mu)$.
Eq.~(\ref{Pilight}) is written for $n_f=3$ active light quarks.
The limit ${\bf q}^2\to 0$ cannot be performed
because there is no scale for light quarks and no PT expression as 
eq.~(\ref{matlep}) is available.

Still small momenta mean IR problems and we want to modify 
only the IR structure of the correlator $\Pi^{light}(\mu^2,{\bf q}^2)$.
It is convenient to modify just the contribution of low energy states 
into the correlator which can be done through 
a dispersion relation.
The dispersion relation reads
\begin{equation}
  \label{dispNew}
\Pi^{light}({\bf q}^2)=
\int\limits_{0}^\infty{\rho^{light}(s)ds\over s+{\bf q}^2}
\end{equation}
where dimensional regularization is understood for $\rho^{light}(s)$.
In fact, eq.~(\ref{dispNew}) can be used for bare quantities
$\Pi^{light}({\bf q}^2)$ and $\rho^{light}(s)$.
The limit ${\bf q}^2\to 0$ in the RHS of eq.~(\ref{dispNew})
is IR-singular and cannot be performed if the PT expression 
for the spectral density $\rho^{light}(s)$ is used. 
Therefore, one should modify the low energy behavior
of the spectrum where PT is not applicable. 
If such a modification is local (has only a finite support in energy
variable $s$ in eq.~(\ref{dispNew})) then it does not affect
any UV properties ($\mu^2$ dependence) 
of $\Pi^{light}(\mu^2,{\bf q}^2)$ which are important for RG.
The low energy modification is inspired by experiment:
at low energies there is a well-pronounced bound state
as a result of strong interaction.
We, therefore, adopt a model of IR modification 
that the high energy tail of the integral in eq.~(\ref{dispNew})
is computed in PT (duality arguments) that retains RG structure of the
result while in the low energy domain there is a contribution of a
single resonance.
An IR modification is performed for contributions of 
$u$, $d$ and $s$ quarks. The massless $u$ and $d$ quarks 
interact with photons through
isotopic combinations $I=1$ ($\rho$-meson channel) and $I=0$ 
($\omega$-meson channel).
For our purposes these two channels are completely degenerate and are
treated simultaneously. 
The $s$-quark contribution is considered separately because 
of its nonvanishing (small) mass $m_s$. 

For a generic light quark correlator in the massless PT approximation
we introduce the IR modification 
\be
\label{lightHadModMod}
\rho^{light}(s)\to \rho_{IRmod}^{light}(s)
=F_R\delta(s-m_R^2)+\rho^{light}(s)\theta(s-s_0)
\ee
where $F_R$, $m_R$ and $s_0$ are IR parameters of the spectrum.
Note that they are not necessarily immediate numbers from experiment.
Substituting the IR modified spectrum (\ref{lightHadModMod})
into eq.~(\ref{dispNew}) one finds
\[
\Pi_{IRmod}^{light}(\mu^2,0)
=\frac{F_R}{m_R^2}+\ln \frac{\mu^2}{s_0}+\frac{5}{3}
+a_s\left(\ln \frac{\mu^2}{s_0}+\frac{55}{12}-4\zeta(3)\right)
\]
\be
\label{lightIRmod}
+a_s^2\left(\frac{9}{8}\ln^2 \frac{\mu^2}{s_0}
+\left(\frac{299}{24}-9\zeta(3)\right)\ln \frac{\mu^2}{s_0}
+\frac{34525}{864}-\frac{715}{18}\zeta(3)+\frac{25}{3}\zeta(5)
-\frac{3 \pi^2}{8}\right)\, .
\ee
Here $a_s\equiv a_s^{(3)}(\mu)$.
We identify $m_R$ with a mass of the lowest resonance
which is the only input giving a scale to the problem.
The IR modifying parameters
$F_R$ and $s_0$ are fixed from duality arguments.

Notice the difference in the ${\cal O}(a_s^2)$ 
order between  eq.~(\ref{Pilight})
and eq.~(\ref{lightIRmod}): in eq.~(\ref{lightIRmod})
there is a new term 
$-3\pi^2/8$. This is so called '$\pi^2$' correction
(e.g. \cite{picorr}). 
It can be rewritten through $\zeta(2)=\pi^2/6$. 

To describe the IR structure of the correlator in representation
given by eq.~(\ref{lightIRmod}) we use OPE with power
corrections that semi-phenomenologically encode 
information about the low energy domain of the spectrum 
through the vacuum condensates of 
local gauge invariant operators \cite{svz}. 
The OPE for the light quark correlator reads
\be
\label{lightOPE}
\Pi^{OPE}(\mu^2,{\bf q}^2)
=\Pi^{light}(\mu^2,{\bf q}^2)
+\frac{\langle {\cal O}_4 \rangle}{{\bf q}^4}
+{\cal O}\left(\frac{\langle{\cal O}_6\rangle}{{\bf q}^6}\right)\, . 
\ee
The quantities $\langle {\cal O}_{4,6} \rangle$ 
give nonPT contributions of dimension-four 
and dimension-six vacuum condensates.
These contributions are UV soft 
(they do not change short distance properties) and
related to the IR modification of the spectrum.
For the purposes of fixing the numerical values of the parameters 
$F_R$ and $s_0$ which describe the 
IR modification of the spectrum
one needs only first two power corrections
$1/{\bf q}^2$ and $1/{\bf q}^4$;
the coefficient of the $1/{\bf q}^2$ correction vanishes
because there are no gauge invariant dimension-two
operators in the massless limit.
Computing the IR modified polarization function and comparing it with 
the OPE 
expression we find finite energy sum rules (FESR)
for fixing the parameters $F_R$ and $s_0$ \cite{FESR1}.
The system of sum rules has the form
\bea
\label{lightsys}
F_R&=&s_0\left\{1+a_s+a_s^2(\beta_0 \ln\frac{\mu^2}{s_0}+k_1+\beta_0)\right\}
+{\cal O}(a_s^3)\, ,\nn \\
F_R m_R^2&=&\frac{s_0^2}{2}
\left\{1+a_s+a_s^2\left(\beta_0 \ln\frac{\mu^2}{s_0}+k_1
+\frac{\beta_0}{2}\right)\right\}
- \langle {\cal O}_4 \rangle + {\cal O}(a_s^3)\, .
\eea
Here $\beta_0=9/4$ and 
\[
k_1=\frac{299}{24}-9\zeta(3)\, .
\]
We treat $\langle {\cal O}_4 \rangle$ as a small correction and
take its coefficient function as a constant
(the total contribution is RG invariant).
Eqs.~(\ref{lightsys}) fix $F_R$ and $s_0$ through $m_R^2$ and 
$\langle {\cal O}_4 \rangle$.
Using higher order terms in OPE 
expansion ($\langle {\cal O}_6 \rangle/{\bf q}^6$) one can avoid
substituting $m_R^2$ from experiment because within the IR
modification given in eq.~(\ref{lightHadModMod}) 
the IR scale is determined by the dimension-six vacuum condensate 
$\langle {\cal O}_6 \rangle$ \cite{FESR1}. 
We do not do that because the primary purpose
is to find the normalization for the EM coupling
and not to describe the spectrum in the low energy domain. 
The use of the experimental value for the
resonance mass $m_R^2$ makes the calculation more precise because
the numerical value for 
$\langle {\cal O}_6 \rangle$ condensate is not known well (cf
ref.~\cite{pivrhodual}).

The leading order solution to eqs.~(\ref{lightsys}) 
(upon neglecting the PT and nonPT corrections) 
is given by the partonic model result
$s_0=2m_R^2$, $F_R=s_0=2m_R^2$, which is rather precise.
This solution has been used for predicting 
masses and residues of the radial excitations
of vector mesons within the local duality approach 
when the experimental spectrum is approximated by 
a sequence of infinitely narrow
resonances \cite{FESR0}.
Such an approximation for the experimental spectrum is
justified by theoretical considerations in 
the large $N_c$ limit \cite{i1nc} and by the exact 
solution for two dimensional QCD \cite{twoqcd}.
For the experimental spectrum of infinitely narrow
resonances the local duality approach 
means averaging over the energy interval around a single resonance
\cite{FESR0}.
It is expected to be less precise than 
the global duality method
in which the average is performed
over the entire spectrum. 
However, within the global duality approach 
only the total contribution of all
hadronic states can be studied while  
the local duality can be used even for first resonances
and can predict characteristics of individual hadronic
states. 

An accurate treatment of eqs.~(\ref{lightsys})
gives the solution
\bea
\label{lightsysSol}
s_0&=&2 m_R^2\left(1+\frac{\beta_0}{2} a_s^2\right)
+\frac{\langle {\cal O}_4 \rangle}{m_R^2}(1-a_s) \, ,\nn \\
\frac{F_R}{m_R^2}&=&2\left\{1+a_s+a_s^2\left(\beta_0 
\ln\frac{\mu^2}{2 m_R^2}+k_1
+\frac{3}{2}\beta_0\right)\right\}
+\frac{\langle {\cal O}_4 \rangle}{m_R^4}\, .
\eea
In the solution given in eq.~(\ref{lightsysSol})
only linear terms in the nonPT correction 
$\langle {\cal O}_4 \rangle$ are retained.
This is well justified numerically. The 
$a_s^2 \langle {\cal O}_4 \rangle$ terms are
neglected because the coefficient function of
the $\langle {\cal O}_4 \rangle$ condensate 
is not known with such precision. In eq.~(\ref{lightIRmod}) 
the scale parameter is $s_0$ while we solve the system 
(\ref{lightsysSol}) in terms
of $m_R$ that we identify with the resonance mass and take from
experiment.
Therefore, we express the PT scale $s_0$ through $m_R$ according to
the solution given in eqs.~(\ref{lightsysSol}). 
The expansion of the log-term in eq.~(\ref{lightIRmod}) reads
\[
\ln\frac{\mu^2}{s_0}=\ln\frac{\mu^2}{2m_R^2}
-\frac{\beta_0}{2}a_s^2
-\frac{\langle {\cal O}_4 \rangle}{2m_R^4}(1-a_s)\, .
\]
With these results one finds an expression 
for the IR modified polarization function of light quarks
at the origin
\[
\Pi_{IRmod}^{light}(\mu^2,0)
=2\left\{1+a_s+a_s^2\left(\beta_0 \ln\frac{\mu^2}{2m_R^2}+k_1
+\frac{3}{2}\beta_0\right)\right\}
+\frac{\langle {\cal O}_4 \rangle}{m_R^4}
\]
\[
+\ln \frac{\mu^2}{2m_R^2}
-\frac{\beta_0}{2}a_s^2
-\frac{\langle {\cal O}_4 \rangle}{2m_R^4}(1-a_s)
+\frac{5}{3}
+a_s\left(\ln \frac{\mu^2}{2m_R^2}-\frac{\langle {\cal O}_4
\rangle}{2m_R^4}
+\frac{55}{12}-4\zeta(3)\right)
\]
\[
+a_s^2\left(\frac{9}{8}\ln^2 \frac{\mu^2}{2 m_R^2}
+\left(\frac{299}{24}-9\zeta(3)\right)\ln \frac{\mu^2}{2 m_R^2}
+\frac{34525}{864}-\frac{715}{18}\zeta(3)+\frac{25}{3}\zeta(5)
-\frac{3 \pi^2}{8}\right)\, .
\]
Here the first line gives the resonance contribution while the rest is the
high energy tail (continuum contribution) which is computed in PT.
Finally, 
\[
\Pi_{IRmod}^{light}(\mu^2,0)
=2\left\{1+a_s+a_s^2\left(\beta_0 \ln\frac{\mu^2}{2m_R^2}+k_1
+\frac{3}{2}\beta_0\right)\right\}
+\frac{\langle {\cal O}_4 \rangle}{2m_R^4}
\]
\[
+\ln \frac{\mu^2}{2m_R^2}
-\frac{\beta_0}{2}a_s^2
+\frac{5}{3}
+a_s\left(\ln \frac{\mu^2}{2m_R^2}
+\frac{55}{12}-4\zeta(3)\right)
\]
\be
\label{lightmat}
+a_s^2\left(\frac{\beta_0}{2}\ln^2 \frac{\mu^2}{2m_R^2}
+k_1 \ln \frac{\mu^2}{2m_R^2}
+\frac{34525}{864}-\frac{715}{18}\zeta(3)+\frac{25}{3}\zeta(5)
-\frac{3 \pi^2}{8}\right)\, .
\ee
Eq.~(\ref{lightmat}) gives $\Pi_{IRmod}^{light}(\mu^2,0)$
as an explicit function of the nonPT scale $m_R$ (to be taken from
experiment)
and theoretical quantities $a_s$ and 
$\langle {\cal O}_4 \rangle$. The choice of the numerical value for 
$a_s$ is discussed in detail later.

The condensate of dimension-four operators for light quarks
is given by
\be
\label{o4numrr}
\langle {\cal O}_4 \rangle
=\frac{\pi^2}{3}\left(1+\frac{7}{6}a_s\right)
\langle\frac{\al_s}{\pi} G^2\rangle 
+ 2\pi^2 \left(1+\frac{1}{3}a_s\right) (m_u+m_d) 
(\langle {\bar u}u \rangle+\langle {\bar d}d \rangle) \, .
\ee
We retain small corrections proportional to the light quark masses
and treat them in the approximation of isotopic symmetry for 
the light quark condensates
$\langle {\bar u}u \rangle = \langle {\bar d}d\rangle$ which is rather
precise for $u$ and $d$ quarks.
The quark condensate part of eq.~(\ref{o4numrr})
is given by the PCAC relation for the $\pi$ meson
\[
(m_u+m_d) \langle {\bar u}u +  {\bar d}d\rangle=-f_\pi^2 m_\pi^2 \, .
\]
Here $f_\pi=133~{\rm MeV}$ is a charged 
pion decay constant and $m_\pi=139.6~{\rm MeV}$
is a charged pion mass. For the standard numerical 
value of the gluon condensate 
$\langle\frac{\al_s}{\pi} G^2\rangle=0.012~{\rm GeV^4}$ \cite{svz}
and $a_s=0.1$ one finds
\be
\label{num040}
\langle {\cal O}_4 \rangle
=
\frac{\pi^2}{3}\left(1+\frac{7}{6}a_s\right)
\langle\frac{\al_s}{\pi} G^2\rangle
-2\pi^2 \left(1+\frac{1}{3}a_s\right)f_\pi^2 m_\pi^2 
= 0.037~{\rm GeV^4}\, .
\ee
For the most important contribution of $u$ and $d$ quarks (the $u$-quark
contribution is enhanced by
factor 4 because of its doubled electric charge in comparison to 
the other light quarks) the relation $s_0=2 m_\rho^2$,
where $m_\rho=768.5~{\rm MeV}$ is a mass of the lowest ($\rho$ meson)
resonance in the non-strange isotopic $I=1$ vector channel,
is rather precise numerically.
The gluon condensate gives a small correction to the basic 
duality relation for light quarks $s_0=2m_R^2$.
Note that we do not identify $F_R$ with the experimental number
available from the analysis of the $\rho$-meson leptonic width but
treat it as an IR modifying parameter which should be fixed from
the consistency requirement with OPE. It is close
numerically to its experimental counterpart 
because it is known that OPE gives rather an accurate description of the
physical spectrum if vacuum condensates 
are included. In the present paper we stick to a theoretical
description of the IR domain 
and use the lowest resonance mass as the only input
for the IR modification. 
The same is true for the $I=0$ channel where the lowest resonance is 
the $\omega$ meson with a mass 
$m_\omega=781.94~{\rm MeV}$. We do not distinguish these two channels.
We consider parameters $F_R$ and $s_0$ as the IR modifiers
fixed theoretically through OPE 
and do not attempt to substitute them from experiment 
(using leptonic decay widths for $F_R$
or the shape of the spectrum for $s_0$). 

Note that the IR parameters of the spectrum $F_R$, $m_R$ and $s_0$  
are $\mu$ independent. It can be seen explicitly from
eqs. (\ref{lightsys}). 

The $n_f=3$ effective theory is valid only
up to ${\bf q}^2\sim m_c^2$ and, formally, there are corrections
of order ${\bf q}^2/m_c^2$ \cite{pivmccorr}. 
However, in the case of current correlators 
these corrections are small \cite{mcch,mclar}.

For the $s$ quark there are also corrections due to $m_s$ which change
slightly the shape of the spectrum and the consistency equations 
for the IR modifiers.
We consider $m_s$ as an additional 
IR modifier which does not affect UV properties 
(renormalization in 
the $\MSsch$ scheme is mass independent) and treat it as a power correction. 
We write OPE for the $s$ quark in the form
\[
\Pi^{OPE,s}(\mu^2,{\bf q}^2)
=\Pi^{light}(\mu^2,{\bf q}^2)
-\frac{6 m_s^2}{{\bf q}^2}
+\frac{\langle {\cal O}_4^s \rangle}{{\bf q}^4}
+{\cal O}\left(\frac{\langle {\cal O}_6 \rangle}{{\bf q}^6}\right)\, .
\]
The system 
of equations for fixing the parameters $F_{Rs}$ and $s_{0s}$ 
reads
\bea
\label{sFESR}
F_{Rs}+6 m_s^2&=&s_{0s}
\left\{1+a_s+a_s^2(\beta_0 \ln\frac{\mu^2}{s_{0s}}+k_1+\beta_0)\right\}
+{\cal O}(a_s^3)\, , \nn \\
F_s m_{Rs}^2&=&\frac{s_{0s}^2}{2}
\left\{1+a_s+a_s^2\left(\beta_0 \ln\frac{\mu^2}{s_{0s}}+k_1
+\frac{\beta_0}{2}\right)\right\}
- \langle {\cal O}_4^s \rangle +{\cal O}(a_s^3)\, .
\eea
Here 
\[
\langle {\cal O}_4^s \rangle 
= \frac{\pi^2}{3}\left(1+\frac{7}{6}a_s\right)
\langle\frac{\al_s}{\pi} G^2\rangle 
+ 8\pi^2 \left(1+\frac{1}{3}a_s\right) m_s \langle {\bar s}s \rangle
\]
is a dimension-four contribution in the strange channel.
One finds the solution to eqs.~(\ref{sFESR})
in the form
\bea
s_{0s}&=&2m_{Rs}^2\left(1+\frac{\beta_0}{2} a_s^2\right)
+\frac{\langle {\cal O}_4^s \rangle}{m_{Rs}^2}(1-a_s)-6m_s^2\, , \nn \\
\frac{F_{Rs}}{m_{Rs}^2}&=&2\left\{1+a_s
+a_s^2\left(\beta_0 \ln\frac{\mu^2}{2m_{Rs}^2}+k_1
+\frac{3}{2}\beta_0\right)\right\}
+\frac{\langle{\cal O}_4^s\rangle}{m_{Rs}^4}-12\frac{m_s^2}{m_{Rs}^2}\, .
\eea
The correction due to $m_s^2$ is not large.
Instead of eq.~(\ref{lightmat}) one has 
\be
\label{strSpol}
\Pi_{IRmod}^{light-s}(\mu^2,0)=\Pi_{IRmod}^{light}(\mu^2,0)
-9\frac{m_s^2}{m_{Rs}^2}
\ee
and $m_{Rs}=m_\varphi$ and $\langle {\cal O}_4^s \rangle$ 
should be used in the first term of eq.~(\ref{strSpol})
instead of 
$m_\rho$ and $\langle {\cal O}_4 \rangle$.
Here $m_\varphi=1019.4~{\rm MeV}$ is a mass of the $\varphi$-meson 
which is the lowest resonance in the strange channel.
A numerical value for $\langle {\cal O}_4^s \rangle$ is obtained as
follows. We use the relation (e.g. \cite{GassLeut})
\[
\frac{2m_s}{m_u+m_d}=25.0
\]
and the phenomenological
result $\langle {\bar s}s \rangle 
= (0.8\pm 0.2)\langle {\bar u}u \rangle$ \cite{gammassuu1}
to find
\be
\label{numss}
m_s \langle {\bar s}s \rangle
= 12.5\cdot 0.8\cdot (m_u+m_d)\langle {\bar u}u \rangle
=-5.0\cdot f_\pi^2 m_\pi^2=-0.0017~{\rm GeV}^4\, .
\ee
One could also use the PCAC relation for the $K$ meson
\[
(m_s+m_u) \langle {\bar s}s+ {\bar u}u \rangle
= -f_K^2 m_K^2+{\cal O}(m_s^2)
\]
with $f_K=1.17f_\pi$ and $m_K=493.7~{\rm MeV}$.
Note that the PCAC relation in the strange channel 
is valid only up to terms of
order $m_s^2$ which are not completely negligible numerically 
compared to the pion case \cite{gammassuu2}.
Therefore, we use the result given in eq.~(\ref{numss}).
For the standard value 
$\langle\frac{\al_s}{\pi} G^2\rangle=0.012~{\rm GeV^4}$ \cite{svz}
and $a_s=0.1$ one finds
\be
\label{num04str}
\langle {\cal O}_4^s \rangle
=
\frac{\pi^2}{3}\left(1+\frac{7}{6}a_s\right)
\langle\frac{\al_s}{\pi} G^2\rangle
+8\pi^2 \left(1+\frac{1}{3}a_s\right)(-5.0 f_\pi^2 m_\pi^2) 
= - 0.0965~{\rm GeV^4}\, .
\ee
The correction due to $m_s \langle \bar{s}s\rangle $ is dominant in 
the dimension four contribution in the strange case.
Because the dimension-four terms represent only small corrections 
to the leading results for the correlators in 
eqs.~(\ref{lightmat},\ref{strSpol}), the precision with
which they are calculated suffices for our purposes.

For the absolute value of $m_s$ to be substituted into $m_s^2$
correction we use the results of recent analyses \cite{msNum}
and take $m_s(M_\tau)=130\pm 27_{exp}\pm 9_{th}~{\rm MeV}$.
For $m_{Rs}=m_\varphi=1019.4~{\rm MeV}$ one finds 
\[
\frac{m_s^2}{m_\varphi^2}= 0.0163
\]
which is a small expansion parameter
that justifies the treatment of $m_s^2$ contribution as a correction.

Note that there are attempts to use constituent masses for the light
quarks and to estimate the polarization functions in the way 
it is done for leptons or heavy quarks.
Besides being {\it ad hoc} (and not supported by experiment)
this IR modification of the light quark correlators 
contradicts OPE and/or
the local duality over the energy interval from the origin to 
$1\div 2 ~{\rm GeV}$.

Thus, eqs.~(\ref{lightIRmod},\ref{lightmat}) represent a
semi-phenomenological subtraction for a light quark
correlator at ${\bf q}^2=0$
based on the IR modification of the spectrum consistent with OPE.
Some mismatch with OPE in orders higher than 
${\cal O}(1/{\bf q}^4)$ which is possible because of simplicity of the IR
modification is neglected. It is justified because we need only
integral characteristics of the spectrum for calculating 
$\Pi_{IRmod}^{light}(\mu^2,0)$ and are not interested in
the 
point-wise behavior of the spectral function $\rho_{IRmod}^{light}(s)$
which is used as an auxiliary quantity in this particular instance. 

\subsection{Heavy quarks}
Matching heavy quarks is straightforward and is similar to that of
leptons. It is done within PT.
For a heavy quark 'q' with the pole mass 
$m_q\gg\Lambda_{\rm QCD}$ one has
\[
\Pi^q(\mu^2,0)=N_c e_q^2\Pi^{heavy}(\mu^2,0)
\]
where $\Pi^{heavy}(\mu^2,0)$ is a generic contribution of a heavy quark
to the polarization function \cite{heavyPi}
\[
\Pi^{heavy}(\mu^2,0)=\ln \frac{\mu^2}{m_q^2}
+e_q^2\frac{\al}{\pi}\left(\frac{45}{16}
+\frac{3}{4}\ln\frac{\mu^2}{m_q^2}\right)
+a_s\left(\frac{15}{4}+\ln\frac{\mu^2}{m_q^2}\right)
\]
\[
+a_s^2\left(\frac{41219}{2592}-\frac{917}{1296}n_l
+\left(4+\frac{4}{3}\ln 2-\frac{2}{3}n_l\right)\zeta(2)
+\frac{607}{144}\zeta(3)
\right.
\]
\be
\label{heavymat}
\left.
\left.
+\left(\frac{437}{36}-\frac{7}{9}n_l\right)\ln \frac{\mu^2}{m_q^2}
+\left(\frac{31}{24}-\frac{1}{12}n_l\right)\ln^2 \frac{\mu^2}{m_q^2}
\right)\right\}
+{\cal O}(\al^2,\al_s^3)\, .
\ee
Here $n_l$ is the number of quarks that are lighter than a heavy one,
$a_s=\al_s/\pi$ is the strong coupling constant in the effective
theory with $n_l+1$ active quarks normalized at the scale $\mu$.
Numbers in eq.~(\ref{heavymat}) 
are given for the pole mass of a heavy quark.
We neglect the (known) EM contribution of order $\al^2$ because it is
smaller than the unknown term of order $\al_s^3$.
Eq.~(\ref{heavymat}) gives a contribution of the corrected partonic model,
i.e. that with a heavy quark loop in the leading approximation.
There is also a contribution of heavy quark loops to the light quark
vacuum polarization function
that should be taken into account
in constructing the effective theory with a decoupled heavy quark.
This contribution is small. It reads \cite{heavyLoop}
\be
\label{heavyLoopMat}
\Pi^{lightheavy}(\mu^2,0)=a_s^2 N_c\left(\sum_{i=1}^{n_l} e_i^2\right)
\left(\frac{295}{1296}-\frac{11}{72}\ln \frac{\mu^2}{m_q^2} 
-\frac{1}{12}\ln^2 \frac{\mu^2}{m_q^2}\right)\, .
\ee
Eqs.~(\ref{heavymat},\ref{heavyLoopMat}) are used for $c$ and $b$ quarks.
Note that these formulas cannot be used for $s$ quark.
Indeed, because of $\al_s$ corrections the PT scale 
in eq.~(\ref{heavymat}) is effectively equal to $m_q$ and
is too small for PT to be applicable
in the case of the strange quark 
since $m_s\sim \Lambda_{\rm QCD}$. 

\section{Low energy normalization: numerics}
In previous sections the necessary contributions due to
fermions have been written down.
We are not going to consider scales larger than $M_Z$
therefore bosonic contributions 
into the EM current and polarization function (namely,
$W$ boson loops) are not 
taken into account. The above equations describe an effective theory
without $W$ bosons which decouple at energies smaller than $M_Z$
and should be added separately for the $Z$-boson peak tests.

A numerical value of the strong coupling at low energies
is rather important for the whole analysis. 
The estimates of the strong coupling numerical value at low scales 
are usually based on the $\tau$ lepton decay data. 
Within a contour resummation technique \cite{Tau,b4}
the value obtained is $\al_s^{(3)}(M_\tau^2)=0.343\pm 0.009_{exp}$.
Within a RG invariant treatment of ref.~\cite{alpha}
a slightly different value
$\al_s^{(3)}(M_\tau^2)=0.318\pm 0.006_{exp}\pm 0.016_{th}$ 
has recently been found.
The uncertainty is due to the experimental 
error and due to truncation of the series
which 
is estimated within an optimistic scenario that higher
order terms are still perturbative (no explicit asymptotic growth).
Note that even for the optimistic scenario with a reduced theoretical
error as compared to the conservative estimates, 
the theoretical error dominates the total uncertainty of the coupling.
Contour improved results include a special resummation procedure 
for treating contributions generated by the running 
which does not necessarily improve results but
definitely changes them in comparison with the finite order 
estimates at the present level of precision. 
The change is still within the error bars
that makes two procedures consistent.
We use the value $\al_s^{(3)}(M_\tau^2)=0.318\pm 0.017$ 
as our basic input for the low energy strong coupling.
The central value $\al_s^{(3)}(M_\tau^2)=0.318$
corresponds to $\al_s^{(5)}(M_Z)=0.118$ when it is run with 
a four-loop 
$\beta$-function and three-loop matching at $m_c$ and $m_b$ thresholds. 

Now one has everything for numerical analysis.
First the value of the running EM coupling constant 
computed in $n_f=4$ effective theory
at $\mu=M_\tau$ that is a convenient normalization point
at the physical mass scale is given. 
Note that the $c$-quark pole mass is
rather close to this value. In fact, 
the recent estimate is $m_c=1.8\pm 0.2~{\rm GeV}$ 
and we take $m_c=M_\tau=1.777~{\rm GeV}$ as a central value,
i.e. $m_c=M_\tau\pm 0.2~{\rm GeV}$. 
Thus, the low energy normalization
value $\bal^{(4)}(M_\tau)$ is computed with 
three active leptons and four active quarks.

For the lepton contribution we use the lepton
masses $M_e=0.5110~{\rm MeV}$, $M_\mu=105.66~{\rm MeV}$,
$M_\tau=1777~{\rm MeV}$ \cite{PDG}.
These values are extremely precise therefore we use them as exact
and assign no errors to them.
We neglect the difference between the running EM coupling $\bal$
and fine structure constant $\al$ in corrections (which results in 
${\cal O}(\al^2)$ shift that is
numerically negligible). We use $\al^{-1}=137.036$. 
According to
eq.~(\ref{matlep}) leptons give 
\[
\Delta^{lept}(M_\tau^2)=\sum_{l=e,\mu,\tau}\Pi^l(M_\tau^2,0)
=\left(1+\frac{3}{4}\frac{\al}{\pi}\right)
\left(\ln \frac{M_\tau^2}{M_e^2}+\ln \frac{M_\tau^2}{M_\mu^2}
+\ln \frac{M_\tau^2}{M_\tau^2}\right)
+\frac{135}{16}\frac{\al}{\pi}
\]
\be
\label{leptonnum}
=21.953+0.058 =  22.011
\ee
where the first number is obtained in the limit $\al=0$.
The $\al$ correction is almost negligible for the normalization 
at the scale $M_\tau$. 
Note that $\tau$ lepton gives no logarithmic contribution at the scale 
$\mu=M_\tau$.

The ${\cal O}(\bal^2)$
correction for the lepton contribution in the $\MSsch$ scheme
is also available \cite{david}. This correction 
is parametrically small and there are no unexpectedly large 
numerical coefficients (in fact, they are also small)
that makes the parametric estimate based on the counting of powers of
$\al$ rather precise.
The sum of contributions of three leptons in ${\cal O}(\bal^2)$ order
is completely negligible and we treat the leptonic
contribution in eqs.~(\ref{matlep},\ref{leptonnum}) as exact.

Light quarks.
From eq.~(\ref{lightmat}) with $m_R=m_\rho$,
and 
$m_{Rs}=m_\varphi$
one finds for the total light quark contribution $\Delta^{uds}(M_\tau^2)$
\[
\Delta^{uds}(M_\tau^2)=\Delta^{u}(M_\tau^2)
+\Delta^{d}(M_\tau^2)+\Delta^{s}(M_\tau^2)
=\Delta^{\rho}(M_\tau^2)+\Delta^{\omega}(M_\tau^2)
+\Delta^{\varphi}(M_\tau^2)
\]
\[
=\frac{4}{3}\Delta^{light}(M_\tau^2)
+\frac{1}{3}\Delta^{light}(M_\tau^2)+\frac{1}{3}\Delta^{light-s}(M_\tau^2)
=\frac{5}{3}\Delta^{light}(M_\tau^2)
+\frac{1}{3}\Delta^{light-s}(M_\tau^2)
\]
\[
=9.13662 + 5.32853 a_s + 24.9086 a_s^2
\]
\[
= 9.13662 + 0.53937 + 0.255214 = 9.9312\, .
\]
Because the calculation is explicit we can give 
the previous result in more detail 
showing all different contributions
\[
\Delta^{uds}(M_\tau^2)= 9.11165 + 0.539367\left(\frac{a_s}{0.101}\right) 
+ 0.2552\left(\frac{a_s}{0.101}\right)^2
\]
\be
\label{udsFin}
+ 0.08865\left(\frac{\langle{\cal O}_4 \rangle}{0.037~{\rm GeV}^4}
\right)
- 0.0488\left(\frac{m_s}{130~{\rm MeV}}\right)^2 
+ 0.0149\left(\frac{\langle{\cal O}_4^s \rangle}{0.0965~{\rm GeV}^4}
\right)\, .
\ee
The IR part of the spectrum (resonances) and  
the partonic quark approximation give a dominant contribution. 
The QCD perturbative 
corrections and power corrections 
due to $m_s$ and $\langle {\cal O}_4^\# \rangle$
condensates are small.
The error is
\be
\delta\Delta^{uds}(M_\tau^2)
=10.5\delta a_s
+0.09\frac{\delta\langle {\cal O}_4 \rangle}{\langle {\cal O}_4 \rangle}
-0.1 \frac{\delta m_s}{m_s} 
-0.015\frac{\delta\langle {\cal O}_4^s \rangle}
{\langle{\cal O}_4^s \rangle}\, .
\ee
Variations $\delta\langle {\cal O}_4^s \rangle$
and $\delta\langle {\cal O}_4 \rangle$
are not completely independent -- both quantities 
contain a variation of the gluon condensate.
Also the error of $a_s$ and that of the gluon condensate 
are correlated (see, for instance, \cite{renorm}).  
For estimating the total error of $\Delta^{uds}(M_\tau^2)$ 
through less correlated quantities
one could rewrite power corrections in eq.~(\ref{udsFin}) in the basis 
of the gluon and strange quark condensates \cite{pivbasis}.
Because the correlation is not well established numerically 
we neglect this effect.
We consider the errors of the strong coupling $a_s$, 
of the gluon condensate for 
$\delta\langle {\cal O}_4 \rangle$, of the strange quark mass $m_s$, 
and of the strange quark condensate $\langle \bar{s}s\rangle$ for 
$\delta \langle {\cal O}_4^s \rangle$ as independent
and use $\delta a_s=0.017/\pi=0.0054$,
$\delta\langle {\cal O}_4 \rangle/{\langle {\cal O}_4\rangle}=1/2$
due to the gluon condensate,
$\delta m_s/{m_s}=0.28$,
$\delta\langle {\cal O}_4^s \rangle/{\langle {\cal O}_4^s \rangle}=1/4$
due to $\langle \bar{s}s\rangle$.
With these (conservative) estimates of 
uncertainties one finds 
\[
\delta\Delta^{uds}(M_\tau^2)=\pm 0.057|_{a_s} 
\pm 0.045|_{\langle {\cal O}_4 \rangle} 
\pm 0.028|_{m_s} \pm 0.004|_{\langle{\cal O}_4^s \rangle}\, .
\]
The dominant error is due to $\delta a_s$.
The gluon condensate gives a sizable 
error because it is enhanced by the charge structure
of light (mainly $u$) quarks
and because its uncertainty is taken to be 
very conservative to compensate for the possible correlation with
$a_s$. The strange channel is suppressed by factor 1/3
in the total sum of light quark contributions 
and its specific features only slightly affect the result:
in the rest it is quite degenerate with $u$ and $d$ channels. 
The total error for the light quark contributions
added in quadrature reads 
\[
\delta\Delta^{uds}(M_\tau^2)=\pm 0.078\, .
\]
The final result for the contribution of light quarks into the 
low energy normalization of the running EM coupling is 
\be
\label{finallight}
\Delta^{uds}(M_\tau^2)=9.9312\pm 0.078\, .
\ee
We retain some additional digits at intermediate stages
just for computational purposes.

For the $c$ quark we use eqs.~(\ref{heavymat},\ref{heavyLoopMat}).
The strong coupling constant in $n_f=4$ effective theory
is found by matching between 
strong coupling constant in $n_f=3$ and 
$n_f=4$ effective theories.

Matching at the pole mass scale $m_P$ for the strong coupling
has the form \cite{matching}
\be
\label{dirmatch3}
a_s^{(n_l)}(m_P^2)=a_s^{(n_l+1)}(m_P^2)\left(1+ C_2 a_s^{(n_l+1)}(m_P^2)^2
+ C_3 a_s^{(n_l+1)}(m_P^2)^3+{\cal O}(a_s^4)\right)
\ee
where
\be 
C_2=-\frac{7}{24}\, ,
\ee
\be 
C_3=-\frac{80507}{27648}\zeta(3)-\frac{2}{9}\zeta(2)(\ln 2+3)
-\frac{58933}{124416}+\frac{n_l}{9}\left(\zeta(2)+\frac{2479}{3456}\right)\, .
\ee
We solve (inverse) eq.~(\ref{dirmatch3}) perturbatively and find
the expression 
\be
a_s^{(n_l+1)}(m_P^2)=a_s^{(n_l)}(m_P^2)\left\{1- C_2 a_s^{(n_l)}(m_P^2)^2
- C_3 a_s^{(n_l)}(m_P^2)^3\right\}
\ee
which is used for determination of the couplings in neighboring effective
theories at their boundary scale that is chosen to be the pole mass of
the heavy quark. 
Matching at $m_c=M_\tau=1.777~{\rm GeV}$ 
(we remind the reader that the numerical value of 
the $c$-quark mass is chosen to be $m_c=M_\tau\pm 0.2~{\rm GeV}$)
with $\al_s^{(3)}(M_\tau^2)=0.318$ gives 
$a_s^{(4)}(m_c^2=M_\tau^2)=0.102$ or 
$\al_s^{(4)}(m_c^2=M_\tau^2)=0.320$. 
This value for the strong coupling is used in eq.~(\ref{heavymat})
for the calculation of the $c$-quark
contribution to the finite renormalization of the EM coupling.
Note that though one computes with $a_s^{(4)}(M_\tau^2)$ it can well be 
identified numerically 
with $a_s^{(3)}(M_\tau^2)$: the change due to matching is
tiny and is much smaller than the error of $a_s^{(3)}(M_\tau^2)$.

According to eqs.~(\ref{heavymat},\ref{heavyLoopMat}) we have 
\[
\Delta^{c}(M_\tau^2)= \Pi^c(\mu^2=M_\tau^2,0)
\]
\be
\label{cqmatch}
=0.00387 + 0.00474+ 0.51001 + 0.32817 = 0.84679
\ee
where the first term is EM contribution, the second one 
is loop contribution
(eq.~(\ref{heavyLoopMat}))
and the last two terms give PT expansion of direct contribution
(eq.~(\ref{heavymat})).
One sees that the EM and loop contributions are much smaller than 
the direct contribution.
Convergence of PT series for the direct contribution is not fast though.

The uncertainty of the $c$-quark contribution
is straightforward to estimate in view of
explicit formulas. The main
error comes from the uncertainty of the $c$-quark mass.
In the next-to-leading order 
one has from eq.~(\ref{heavymat})
\be
\label{cquarkunce}
\delta \Delta^c(M_\tau^2)=-\frac{4}{3}(1+a_s)\frac{2\delta m_c}{m_c}
=-\frac{8}{3}(1+a_s)\frac{\delta m_c}{m_c}=\pm 0.330
\ee
for $m_c=M_\tau\pm 0.2~{\rm GeV}$
and $a_s=0.1$.
This is a very large uncertainty.
The contribution $\Delta^c(M_\tau^2)$
in eq.~(\ref{cqmatch})
is small because the $c$ quark almost decouples but the uncertainty
of $\Delta^c(M_\tau^2)$ is large. The uncertainty 
is mainly given by the variation of $\ln(M_\tau^2/m_c^2)$
in eq.~(\ref{heavymat}) 
and is independent of an absolute value of the contribution.
For the central value $m_c=M_\tau$ one would find a vanishing
contribution in the leading order but its uncertainty would 
almost stay unchanged and equal to 0.330.
Also the $c$-quark mass is not very large and the convergence of the
PT expansion in eqs.~(\ref{heavymat},\ref{cqmatch}) is slow.

Note that for estimating 
the $c$-quark contribution we do not take into account 
the $a_s$ uncertainty.
The reason is that uncertainties of the quark mass and of $a_s$ are
strongly correlated.
Indeed, to the leading order one can find from eq.~(\ref{heavymat})
\be
\label{asuncerpole}
\delta \Pi^{heavy}(M_\tau^2,0)=\frac{15}{4}\delta a_s
\ee
for independent variation of $a_s$. However, eq.~(\ref{heavymat})
can be rewritten in
terms of the running mass ${\bar m}_c(\mu^2)$. To the first order in $a_s$
the relation between masses reads
\be
\label{mpole2mrun}
m_c={\bar m}_c(\mu^2)\left\{1
+a_s(\mu^2)\left(\ln\frac{\mu^2}{{\bar
m}_c^2(\mu^2)}+\frac{4}{3}\right)\right\}
\ee
that leads to the change in eq.~(\ref{heavymat})
\[
\ln \frac{\mu^2}{m_c^2}
+a_s\left(\frac{15}{4}+\ln\frac{\mu^2}{m_c^2}\right)
\quad \to \quad
\ln \frac{\mu^2}{{\bar m}_c^2(\mu^2)}
+a_s\left(\frac{13}{12}-\ln\frac{\mu^2}{{\bar m}_c^2(\mu^2)}\right)\, .
\]
The NLO result for the polarization function in terms of the running
mass now reads
\[
\Pi_{runmass}^{heavy}(\mu^2,0)=\ln \frac{\mu^2}{{\bar m}_c^2(\mu^2)}
+a_s\left(\frac{13}{12}-\ln \frac{\mu^2}{{\bar m}_c^2(\mu^2)}\right)\, .
\]
This result leads to the uncertainty
\[
\delta \Pi_{runmass}^{heavy}(M_\tau^2,0)
=\left(\frac{13}{12}
-\ln \frac{M_\tau^2}{{\bar m}_c^2(\mu^2)}\right)\delta a_s
\]
which is by factor 4 smaller 
numerically than the previous result eq.~(\ref{asuncerpole}).
The rest of the uncertainty is now in the relation between 
the pole and
running mass given in eq.~(\ref{mpole2mrun})
that represents 
a regular change of variables in the finite order PT
and is under rather a strict
control. We work with the pole mass and 
assume that the uncertainty of the polarization function
at the origin is saturated by the uncertainty of the pole mass. 
It is also assumed that
the uncertainty of the pole mass is estimated 
such that it includes an uncertainty of $a_s$.  

The total finite renormalization between the fine structure 
constant and the $\MSsch$-scheme EM coupling at $M_\tau$ is given by 
\[
\Delta^{(4)}(M_\tau^2)=\Delta^{lept}(M_\tau^2)+\Delta^{uds}(M_\tau^2)
+\Delta^{c}(M_\tau^2)
\]
\be
\label{lightMatch}
=22.0109+9.9312+0.84679= 32.7889
\ee
that leads to  
\be
\label{finalbal400}
\frac{3\pi}{\bal^{(4)}(M_\tau^2)}
=\frac{3\pi}{\al}-\Delta^{(4)}(M_\tau^2)
=\frac{3\pi}{\al}-32.7889\, .
\ee
The low energy normalization value for the EM coupling 
in the $\MSsch$ scheme reads 
\be
\frac{1}{\bal^{(4)}(M_\tau^2)}=133.557\, .
\ee
We now consider the uncertainty of this central result.
The lepton contributions are treated as exact so the number 
from eq.~(\ref{leptonnum}) has no errors.
The errors due to light quarks are given in eq.~(\ref{finallight}).
Note that one could reduce the sensitivity 
of $\Delta^{uds}(M_\tau^2)$ to $a_s$, 
the error of which dominates the total error
in eq.~(\ref{finallight}),
by taking $F_R$ from experiment through the leptonic
decay width of the $\rho$ meson (and of the
$\omega$ and $\varphi$ mesons in other light quark channels). 
Then the resonance contribution  
$F_R/m_R^2$ does not depend on $a_s$.
The first duality
relation fixes $s_0$ immediately using the fact that power
corrections
of order $1/q^2$ are absent in the OPE.
However, this procedure introduces an experimental 
error due to an uncertainty in the $\rho$-meson leptonic decay width 
\[
\Gamma_{ee}^\rho=6.77\pm 0.32~{\rm keV}\, .
\]
This uncertainty leads to almost the same
error for the final quantity $\Delta^{uds}(M_\tau^2)$
as the uncertainty in $a_s$.
It seems to be natural.
Indeed, the 
strong coupling at low energies is extracted from the $\tau$ data in which
the $\rho$-meson contribution constitutes an essential part.
This example shows how the coupling constant encodes 
information on the experimental data.
Another point about using $\Gamma_{ee}^\rho$
for the lowest resonance contribution
is that the consistency with OPE is less strict for such a procedure
(no dimension-four operators participate). 
Still having in mind the possibility of further improvement through 
experiment we consider our estimate of the error given in 
eq.~(\ref{finallight}) as rather conservative.

The uncertainty of the $c$-quark contribution is given in
eq.~(\ref{cquarkunce}).
Collecting all together
one finds the final prediction
\be
\label{finaldel4}
\Delta^{(4)}(M_\tau^2)
=32.7889\pm 0.078_{light}\pm 0.330_{c}
\ee
and 
\be
\label{finalbal4}
\frac{3\pi}{\bal^{(4)}(M_\tau^2)}
=\frac{3\pi}{\al}-\Delta^{(4)}(M_\tau^2)
=\frac{3\pi}{\al}-(32.7889 \pm 0.078_{light}\pm 0.330_{c})\, .
\ee
Eq.~(\ref{finalbal4}) is the main result
for the low energy normalization of the running EM coupling. 
For the coupling itself it reads
\be
\label{finiatauB}
\frac{1}{\bal^{(4)}(M_\tau^2)}=133.557\pm  0.0083_{light}
\pm 0.0350_{c}
\ee
and 
\[
{\bal^{(4)}(M_\tau^2)}=1.0261\al.
\]
This value $\bal^{(4)}(M_\tau^2)$ (or equivalently $\Delta^{(4)}(M_\tau^2)$)
represents the boundary (initial) condition for the running.  
With this value known the EM coupling can be run to other scales. 
The final goal is $M_Z=91.187~{\rm GeV}$ where high precision tests of
SM are done. As will be seen later the running itself is very precise 
numerically and the main uncertainty in the running EM coupling 
at larger scales is due to
the boundary condition eq.~(\ref{finiatauB}). The boundary condition
eq.~(\ref{finiatauB}) has rather a big uncertainty because of the
error of the $c$-quark mass mainly. 
The uncertainty due to the light quark contribution 
is reasonably small. It is dominated by the error in $a_s(M_\tau)$ 
which is mainly theoretical, i.e. related to the truncation of PT
series used for describing the $\tau$-lepton decay data.
The uncertainty in $a_s(M_\tau)$ can be reduced 
if some other sources for its determination are used
in addition to the $\tau$ system.
Reducing the $c$-quark pole mass uncertainty 
requires more accurate treatment of the threshold region of $c\bar c$
production which is rather a challenging problem in QCD.

\section{RG evolution in MS scheme}
With the boundary value known at sufficiently large scale
one can perturbatively run the EM coupling to larger scales. 
The final goal is the determination of the numerical 
value for the EM coupling at $M_Z$ where high precision 
tests of the standard model are performed.
The running itself (as a functional) is extremely precise
because $\beta$-functions are very well known.
The precision of running is affected by the initial value of $a_s$ 
which is chosen to be $a_s^{(3)}(M_\tau^2)$ and by the $b$-quark mass $m_b$.
We discuss them in detail later.

\subsection{Basic relations for RG evolution}
For the evolution between the $\tau$-lepton mass 
$M_\tau=1.777~{\rm GeV}$ (numerically
$m_c=M_\tau$) and $M_Z=91.187~{\rm GeV}$
the number of active quarks is either 4 or 5 and only one
threshold at $m_b$ is encountered. 
The evolution equation (running) is written in the form 
\[
\mu^2\frac{d}{d\mu^2}\left(\frac{3\pi}{\bal(\mu^2)}\right)
=3\left(1+\frac{3}{4}\frac{\bal}{\pi}\right)
+\left(\frac{10}{3}+\frac{1}{3}\theta_b\right)
+\left(\frac{17}{18}+\frac{1}{36}\theta_b\right)\frac{\bal}{\pi}
\]
\be
\label{runQCD}
-\left(\frac{34}{27}+\frac{1}{27}\theta_b\right)
\frac{\bal}{4\pi}a_s
+a_s h^{\rm QCD}(a_s)\, .
\ee
Here $\theta_b$ is a parameter for the $b$-quark presence, $n_f=4+\theta_b$.
From $M_\tau$ to $m_b$ one has $n_f=4$ and $\theta_b=0$ 
while from $m_b$ to $M_Z$ one has $n_f=5$ and $\theta_b=1$. 
In eq.~(\ref{runQCD})
the strong coupling $a_s(\mu^2)$ obeys RGE
\begin{equation}
\label{RGeqStr}
\mu^2\frac{d}{d\mu^2}a_s(\mu^2)=\beta(a_s(\mu^2))
+a_s^2\frac{\bal}{8\pi}\left(\sum_{q}e_q^2\right)
\end{equation}
with 
\be
\label{betaStr}
\beta(a_s)=-a_s^2(\beta_0 +\beta_1 a_s+\beta_2 a_s^2
+\beta_3 a_s^3)+{\cal O}(a_s^6)
\ee
being the strong interaction $\beta$-function.
In QCD one has
\[
h^{\rm QCD}(a_s)=\left(\frac{10}{3}+\frac{1}{3}\theta_b\right)
\left\{1+a_s\left(\frac{287}{144}-\frac{11}{72}\theta_b\right)
\right.
\]
\[
\left.
+a_s^2\left(\frac{38551}{15552}-\frac{7595}{7776}\theta_b
-\frac{77}{3888}\theta_b^2-\frac{55}{54}\zeta(3)(1+\theta_b)
\right)   
\right\}
\]
\be
\label{hqcdFu}
+a_s^2\left(\frac{2}{3}-\frac{1}{3}\theta_b\right)^2
\left(\frac{55}{72}-\frac{5}{3}\zeta(3)
\right)   
\ee
where the first two lines give the 'direct'
contribution and the third line gives light-by-light contribution 
which is written separately
because of its different color structure. 
This result is obtained from the photon renormalization constant given in
\cite{surgu}
and explicitly written in \cite{heavyLoop,chetgluino}.
It was used in ref.~\cite{erler} for calculation of the 
evolution of the EM coupling constant.
Numerically, one finds
\[
h^{\rm QCD}(a_s)=\left(\frac{10}{3}+\frac{1}{3}\theta_b\right)
\left(1+a_s(1.993 - 0.153\theta_b)
+a_s^2(1.26 - 2.20\theta_b  - 0.02\theta_b^2)\right)
\]
\be
\label{numh}
+a_s^2(-0.55 + 0.55\theta_b - 0.14\theta_b^2)\, .
\ee
Coefficients of the EM $\beta$-function 
in eqs.~(\ref{runQCD},\ref{hqcdFu},\ref{numh}) are very small
that makes the convergence of PT series for the evolution very fast.
Eqs.~(\ref{runQCD},\ref{RGeqStr}) should be solved simultaneously.
However, the EM coupling $\bal(\mu)$
is small, therefore, we neglect its running
in corrections and substitute there the value numerically
equal to the fine structure constant $\al$.
Then one has to integrate the trajectory of the strong
coupling $a_s(\mu)$ which is given by the solution to RGE (\ref{RGeqStr}).
The $\al$ correction in the strong coupling 
$\beta$-function is numerically of order $a_s^2$ and formally 
should be retained if $a_s^4$ terms in the $\beta$-function 
are retained. However, the main contribution to the running is 
given by the partonic part of the EM $\beta$-function
in eq.~(\ref{runQCD}), i.e. independent of both EM and strong
couplings. Other terms give only small corrections.
As for practical calculations, one can do everything numerically,  
however,
it happens that the two-loop running gives almost the same result as the exact
treatment. With the two-loop accuracy the integration can be done analytically
in a simple form.
Indeed, for $\beta(a_s)=-\beta_0 a_s^2-\beta_1 a_s^3$ 
one finds
\bea
\label{traint}
\int\limits_{\mu_1^2}^{\mu_2^2} a_s(\xi) d\ln \xi
&=&\frac{1}{\beta_0}\ln\left(\frac{\beta_0/a_s(\mu_2^2)+\beta_1}
{\beta_0/a_s(\mu_1^2)+\beta_1}\right), \nn \\
\int_{\mu_1^2}^{\mu_2^2} a_s(\xi)^2\frac{d\xi}{\xi}
&=&-\frac{1}{\beta_1}\ln\left(\frac{\beta_0+\beta_1 a_s(\mu_2^2)}
{\beta_0+\beta_1 a_s(\mu_1^2)}\right)
\eea
where the NLO solution for the running coupling $a_s(\mu)$
is given by
\be 
\label{intb2}
\ln \left(\frac{\mu^2}{\Lambda^2}\right) 
=\Phi(a_s) = \int^{a_s}  \frac{d\xi}{-\xi^2(\beta_0+\beta_1 \xi)}
= \frac{1}{a_s \beta_0} 
       + \frac{\beta_1}{\beta_0^2} 
\ln \left( \frac{a_s \beta_0^2}{\beta_0 + a_s \beta_1} \right) \, .
\ee
In NNLO it is also possible to perform integration explicitly 
but results are too awkward to present here.
In fact, the NLO integration as given in eqs.~(\ref{traint},\ref{intb2})
is rather precise numerically
and can be used for preliminary estimates.
We, however, avoid any approximation of this sort (cf
ref.~\cite{erler}) and 
give numbers for a direct numerical treatment
of RG equations~(\ref{runQCD},\ref{RGeqStr}) 
with the four-loop strong coupling $\beta$-function
from eq.~(\ref{betaStr})
and $h^{\rm QCD}$-function from eq.~(\ref{hqcdFu}).

The solution to RGE can be used for 
the range of $\mu$ where the corresponding 
effective theory (with a given number of active leptons and quarks)
is valid. Because decoupling is not automatic
one should explicitly take into account thresholds.

\subsection{Running to $m_b$}
The first scale of interest is $m_b$
where there is an important physics due to $b\bar b$ production near
threshold and accurate data on $\Upsilon$ resonances (note that the 
real threshold energy is,
in fact, $2 m_b$ but the matching is defined at $m_b$). We use
$m_b=4.8\pm 0.2~{\rm GeV}$ as determined in ref.~\cite{bbmass}.

In the approximation 
when the EM coupling is taken to be constant in the correction,
the contribution of leptons is given by 
\be
\Delta_{\tau b}^{lept}(m_b^2)
=3(1+\frac{3}{4}\frac{\al}{\pi})\ln \frac{m_b^2}{M_\tau^2}
=3(1+\frac{3}{4}\frac{\al}{\pi})\cdot 1.98738=5.9725\, .
\ee
The hadronic part is more involved.
In the energy range from $M_\tau$ to $m_b$
the number of active quarks is 4 or $\theta_b=0$.
The partonic part of quark contribution reads
\be
\Delta_{\tau b}^{(0)}(m_b^2)
=N_c \sum_{q}e_q^2\left(1+e_q^2\frac{3}{4}\frac{\bal}{\pi}\right)
\ln \frac{m_b^2}{M_\tau^2} 
=\left(\frac{10}{3}
+\frac{17}{18}\frac{\al}{\pi}\right)\ln \frac{m_b^2}{M_\tau^2}= 6.62896
\ee
where we use ${\bal}=\al$.
The result is independent of strong coupling constant 
(the parton model result without real QCD interaction).
The quark part beyond the partonic approximation requires 
integration of the evolution trajectory of the strong coupling
in $n_f=4$ effective theory.
The initial value of the strong coupling
is $a_s^{(4)}(M_\tau^2)=0.102001$
as was obtained from matching at $M_\tau^2$ 
for the $c$-quark contribution. In NLO one finds
still a sizable contribution
\be
\label{fortable4}
\Delta_{\tau b}^{(1)}(m_b^2)
=(\frac{10}{3}-\frac{17}{54}\frac{\al}{\pi})I_{\tau b}^{(1)}=0.54878\, .
\ee
The NNLO contribution proportional to $a_s^2$ in eq.~(\ref{runQCD})
\be
\Delta_{\tau b}^{(2)}(m_b^2)
=\frac{10}{3}\frac{287}{144}I_{\tau b}^{(2)}=0.091848
\ee
and the NNNLO contribution proportional to $a_s^3$ in eq.~(\ref{runQCD})
\be
\Delta_{\tau b}^{(2)}(m_b^2)=\left(\frac{200675}{23328}
-\frac{335}{81}\zeta(3)\right)
I_{\tau b}^{(3)}=3.63085 I_{cb}^{(3)}=0.0042699
\ee
give only small corrections.
Here 
\[
I_{\tau b}^{(n)}=\int_{M_\tau^2}^{m_b^2}(a_s^{(4)}(s))^n\frac{ds}{s}\, .
\]
A total correction to the parton model result (i.e. QCD contribution)
\be
\Delta_{\tau b}^{(hadcor)}(m_b^2)=
\Delta_{\tau b}^{(1)}(m_b^2)+\Delta_{\tau b}^{(2)}(m_b^2)
+\Delta_{\tau b}^{(3)}(m_b^2)
=0.644899
\ee
is much smaller than the leading partonic result
$\Delta_{\tau b}^{(0)}(m_b^2)$.
For the EM coupling at $m_b$ one finds
\bea
\label{finalbal4mb00}
\frac{3\pi}{\bal^{(4)}(m_b^2)}
&=&\frac{3\pi}{\bal^{(4)}(M_\tau^2)}-(5.9725+6.62896+0.644899)\nn \\
&=&\frac{3\pi}{\bal^{(4)}(M_\tau^2)}-13.2464\, .
\eea
Lepton and parton contributions dominate. 
Collecting all together one finds
\[
\frac{3\pi}{\bal^{(4)}(m_b^2)}
=\frac{3\pi}{\bal^{(4)}(M_\tau^2)}-13.2464
\]
\be
=\frac{3\pi}{\al}-(32.7889 + 13.2464)=\frac{3\pi}{\al}-46.0353 \, .
\ee
And finally  
\be
\frac{1}{\bal^{(4)}(m_b^2)}=132.152
\ee
or 
\be
\bal^{(4)}(m_b^2)=1.037\al \, .
\ee
This number can be used for the $\Upsilon$-resonance physics analysis.

Because decoupling is not explicit in mass-independent
renormalization schemes there is another EM coupling parameter 
related to the scale $m_b$. Upon changing the number of active quarks to 
$n_f=5$ one obtains
\be
\label{matchem4to500}
\frac{3\pi}{\bal^{(4)}(m_b^2)}
=\frac{3\pi}{\bal^{(5)}(m_b^2)}+\Pi^{bfull}(\mu^2=m_b^2,0)\, .
\ee
The polarization function $\Pi^{bfull}(\mu^2=m_b^2,0)$ 
which gives the corresponding shift for the EM constant 
is written in terms of the effective 
strong coupling constant $a_s^{(5)}(m_b^2)$.
A numerical value for 
$a_s^{(5)}(m_b^2)$ is obtained through 
matching the strong coupling at the scale $m_b$. 
The running of the coupling
$a_s^{(4)}(M_\tau^2)=0.102$ to $m_b=4.8~{\rm GeV}$
gives 
$a_s^{(4)}(m_b^2)=0.06851$ ($\al_s^{(4)}(m_b^2)= 0.2152$).
Then matching at $m_b$ results in
$a_s^{(5)}(m_b^2)=0.06869$.
With this number 
the result of matching for the EM constant is
\[
\Delta^{b}(m_b^2)=\Pi^{bfull}(\mu^2=m_b^2,0)
=\Pi^{bdir}(\mu^2=m_b^2,0)+\Pi^{bloop}(\mu^2=m_b^2,0)
\]
\be
= 0.00024_{EM} + 0.00358_{loop}+ 0.08587 + 0.03437 =0.1241\, .
\ee
The EM contribution is totally negligible. The loop 
contribution is rather small.
The PT convergence of the direct contribution
is not fast and is similar to the $c$-quark case.
One has
\be
\label{mbmatchfull}
\Delta^{b}(m_b^2)=0.1241\, .
\ee
Finally, the EM couplings of $n_f=4$ and $n_f=5$ 
effective theories in the vicinity of $m_b$ are related by
\be
\label{finalbal5mb}
\frac{3\pi}{\bal^{(5)}(m_b^2)}=
\frac{3\pi}{\bal^{(4)}(m_b^2)}-\Delta^{b}(m_b^2)
=\frac{3\pi}{\bal^{(4)}(m_b^2)}-0.1241\, .
\ee
Explicitly one finds
\[
\bal^{(5)}(m_b^2)= \frac{1}{132.138}=1.0001\bal^{(4)}(m_b^2)\, .
\]
This difference can be safely neglected
in applications for $\Upsilon$-resonance physics.

The uncertainty due to $m_b$ is tiny.
Indeed, the error in the $b$-quark
mass leads to the uncertainty
\be
\label{mbmatchUnce}
\delta \Delta^{b}(m_b^2)
=-\frac{1}{3}(1+a_s^{(5)}(m_b^2))\frac{2\delta m_b}{m_b}
=-\frac{2}{3}(1+a_s^{(5)}(m_b^2))\frac{\delta m_b}{m_b}=\pm 0.030\, .
\ee
There are two reasons for such a smallness in comparison to 
the $c$-quark case: 
the electric charge of $b$ quark $|e_b|$ is two times smaller 
than $|e_c|$ and the relative
uncertainty of the $b$-quark mass $m_b$ ($\delta m_b/m_b$) 
is much smaller than that of the $c$-quark mass.
Note that because the contribution $\Delta^{b}(m_b^2)$
itself is small the relative
uncertainty $\delta \Delta^{b}(m_b^2)/\Delta^{b}(m_b^2)$
is huge. However, one cannot use it here. Even for 
$\Delta^{b}(m_b^2)=0$ the uncertainty 
$\delta \Delta^{b}(m_b^2)$ is basically $0.030$.

\subsection{Running from $m_b$ to  $M_Z$.}
In this subsection we describe 
the evolution of the EM coupling constant $\bal^{(5)}(m_b^2)$
from $m_b=4.8~{\rm GeV}$ to $M_Z=91.187~{\rm GeV}$.
Various contributions according to eqs.~(\ref{runQCD},\ref{hqcdFu}) 
are:

\noindent a) Leptonic contribution
\be
\Delta_{bZ}^{lept}=
3\left(1+\frac{3}{4}\frac{\al}{\pi}\right)
\ln \frac{M_Z^2}{m_b^2}=17.6966\, ;
\ee
b) Leading quark partonic, $a_s$-independent, contribution
\be
\label{bZdelta0}
\Delta_{bZ}^{(0)}=\left(\frac{11}{3}
+\frac{35}{36}\frac{\al}{\pi}\right)\ln\frac{M_Z^2}{m_b^2} 
= 21.6048\, ;
\ee
c) the NLO contribution with 
$a_s^{(5)}(m_b^2)=0.068694$
as initial value for the evolution trajectory
\be
\Delta_{bZ}^{(1)}=\left(\frac{11}{3}-\frac{35}{108}\frac{\al}{\pi}\right)
I_{bZ}^{(1)}=1.0780\, ;
\ee
d) the NNLO contribution proportional to $a_s^2$
\be
\Delta_{bZ}^{(2)}=\frac{11}{3}\frac{265}{144}I_{bZ}^{(2)}= 0.10213\, ;
\ee
e) the NNNLO contribution proportional to $a_s^3$
\be
\Delta_{bZ}^{(3)}=\left(\frac{257543}{46656}
-\frac{620}{81}\zeta(3)\right)I_{bZ}^{(3)}=-3.68089 I_{bZ}^{(3)}
=-0.002954\, .
\ee
Here
\[
I_{bZ}^{(n)}=\int_{m_b^2}^{M_Z^2}(a_s^{(5)}(s))^n\frac{ds}{s}\, .
\]
A total QCD correction to the partonic result 
\be
\Delta_{bZ}^{(hadcor)}= \Delta_{bZ}^{(1)}
+\Delta_{bZ}^{(2)}+\Delta_{bZ}^{(3)}=1.1772
\ee
is small compared to 
the leading quark partonic, $a_s$-independent, contribution
given in eq.~(\ref{bZdelta0}).
The total effect of running on the interval from $m_b$ to $M_Z$
\be
\label{mb2mZrun}
\Delta_{bZ}^{lept}+\Delta_{bZ}^{(0)}
+\Delta_{bZ}^{(hadcor)}= 17.697+21.6048+1.1772= 40.479
\ee
is dominated by leptons and by the quark partonic contribution.
We find the EM coupling at $M_Z$
expressed through the EM coupling at $m_b$ in the form
\be
\label{finalbal5mZ}
\frac{3\pi}{\bal^{(5)}(M_Z^2)}
=\frac{3\pi}{\bal^{(5)}(m_b^2)} - 40.479\, .
\ee
This equation gives the relation between the running EM
couplings necessary for applications in $b$- and $Z$-physics.

Collecting together 
eqs.~(\ref{finalbal400},\ref{finalbal4mb00},\ref{mbmatchfull},\ref{finalbal5mZ}) 
we find the absolute value of the running EM coupling at $M_Z$
expressed through the fine structure constant $\al$ 
\[
\frac{3\pi}{\bal^{(5)}(M_Z^2)}=\frac{3\pi}{\al}
-32.7889(match~M_\tau)-13.2464(run~m_c2m_b)
\]
\be
\label{finalbal5mZcol}
-0.1241(match~m_b)-40.479(run~m_b2M_Z)
=\frac{3\pi}{\al}-86.6384
\ee
and 
\be
\frac{1}{\bal^{(5)}(M_Z^2)}=\frac{1}{\al}
-86.6384/(3\pi)=137.036-9.1926=127.843 \, .
\ee
The result can be written as a relation between the running EM
coupling
and the fine structure constant
\be
\bal^{(5)}(M_Z^2)=1.0719 \al \, .
\ee
This number can be used for the $Z$-boson peak analysis.

\section{Summary of results}
In this section we give a brief summary of the calculation
paying attention to uncertainties
of the results. 

The uncertainty for the low energy normalization value at the $\tau$ mass
is given in eqs.~(\ref{finaldel4},\ref{finalbal4},\ref{finiatauB}).
It is largely dominated by
the uncertainty due to the $c$-quark contribution.
Adding uncertainties 
\[
\delta \Delta^{(4)}(M_\tau^2)=\pm 0. 078_{light}\pm 0.330_c \, .
\]
in quadrature one finds
\[
\delta \Delta^{(4)}(M_\tau^2)=\pm 0.339
\]
and 
\[
0.339/(3\pi)=0.036.
\]
Finally one obtains for the low energy normalization value
the result of the form
\be
\frac{1}{\bal^{(4)}(M_\tau^2)}=133.557\pm 0.036\, .
\ee
The total error is dominated by the uncertainty due to the $c$-quark matching
contribution which is mainly given by the 
uncertainty of the $c$-quark mass. 

\begin{table}
\[ 
\begin{array}{||c||c|c|c||c||}
\hline \hline
{\rm Range}   &(0-M_\tau)&M_\tau - m_b& m_b-M_Z & total   \nn \\ \hline
\Delta^{lept} & 22.011   &   5.973    &17.697   & 45.681   \\
\hline \hline 
\end{array}
\]
\caption{\label{Tab1}Leptonic contributions.}
\end{table}

\begin{table}
\[ 
\begin{array}{||c||c|c|c||}
\hline \hline
{\rm Matching}  &\Delta^{uds}(M_\tau) &\Delta^{c}(M_\tau )
&\Delta^{b}(m_b)    \nn \\ \hline
{\rm Value } &  9.9312  \pm 0.078   &  0.8468  \pm 0.330   
& 0.1241 \pm 0.030   \\
\hline \hline 
\end{array}
\]
\caption{\label{Tab2}Matching at different scales.}
\end{table}

For the scales $m_b$ and $M_Z$, the errors due to running, which are
basically because
of the uncertainty of the coupling constant $a_s$, should be included.
The running itself is precise because $\beta$-functions in  
eqs.~(\ref{runQCD},\ref{RGeqStr}) are computed up to high order of 
PT and the coupling constant $a_s$ is rather well known.
The dominant contribution comes from leptons and 
partonic quarks and is independent of the genuine QCD interaction
(see Tables \ref{Tab1},\ref{Tab3}).
The EM terms give a tiny correction.
In Table \ref{Tab4} the quantity 
$\int_{M_\tau^2}^{m_b^2}a_s(s)\frac{ds}{s}$
with running for $a_s(s)$ in different orders and with or without EM
contribution to the strong $\beta$-function is presented.
The inclusion of EM terms 
slows down the decrease of the strong coupling and the integrals are
slightly larger; still it is completely negligible numerically. 
In the leading order we have the uncertainty in integrals due to 
errors of the initial value of the strong coupling
\be
\label{runErrGen}
\delta I_{ab}^{(1)}=\frac{L_{ab}}{1+a_s \beta_0 L_{ab}}\delta a_s
\ee
with $L_{ab}=\ln \mu_b^2/ \mu_a^2$.
This equation suffices for error estimates
of the QCD contribution into running. NNLO and NNNLO give only small
corrections. One can find uncertainty of the running by varying the
initial value of $a_s$. The numerical results are close to the
estimate
given in eq.~(\ref{runErrGen}). 
Eq.~(\ref{runErrGen}), however, has an advantage of
being analytical and simple that makes the error 
evaluation more transparent. 
 
\begin{table}
\[ 
\begin{array}{||c||c|c||c||}
\hline \hline
{\rm Power}          &M_\tau - m_b & m_b-M_Z  & {\rm total}  \nn \\ \hline
\Delta^{(0)}\sim a_s^0 & 6.6290       &  21.605    &28.234    \\
\Delta^{(1)}\sim a_s^1 & 0.5488      &  1.078     &1.627    \\
\Delta^{(2)}\sim a_s^2 & 0.0918       &  0.102      &0.194    \\
\Delta^{(3)}\sim a_s^3 & 0.0043     &  -0.003    &0.001    \\\hline\hline
{\rm total~sum}           & 7.2739       &  22.782     &30.056   \\\hline
\Delta^{(hadcor)}   & 0.6449       &   1.177     &1.822    \\
\hline \hline 
\end{array}
\]
\caption{\label{Tab3}Running of powers of $a_s$.}
\end{table}

At $m_b$ (and $M_Z$) errors due to running and due to matching 
the light quark contribution at
$M_\tau$ are not independent:
both are determined mainly by the uncertainty in $a_s$.
Therefore these errors should be added linearly.

For the interval from $M_\tau$ to $m_b$ one has from eq.~(\ref{runErrGen})
\[
\delta\Delta^{hadcor}_{\tau b}|_{a_s}=4.56\delta a_s=0.025
\]
and the total error (with linearly added errors for matching 
the light quark contribution at $M_\tau$ and running) is 
\[
\delta\Delta^{(5)}(m_b^2)=\pm (0.078+0.025)_{light+run}\pm 0.330_c
\pm 0.030_b 
\]
\[
=\pm 0.103_{light+run}\pm 0.330_c \pm 0.030_b \, .
\]
Adding independent errors in quadrature one has
\[
\delta\Delta^{(5)}(m_b^2)=\pm 0.347
\]
and 
\[
 0.347/(3\pi)= 0.0368\, .
\]
Finally, one finds the uncertainty for the EM coupling at $m_b$
\[
\frac{1}{\bal^{(4)}(m_b)}\approx
\frac{1}{\bal^{(5)}(m_b)}=\frac{1}{\al}-\Delta^{(5)}(m_b^2)
\]
\be
=137.036-(4.89766\pm 0.0368)= 132.138\pm 0.0368\, .
\ee

\begin{table}
\[ 
\begin{array}{||c||c|c||}
\hline \hline
{\rm Order}    &I^{(1)}_{\tau b}~with~{\rm EM} 
&I^{(1)}_{\tau b}~without~{\rm EM}  \nn \\ \hline
LO             & 0.169104     &0.169100         \\
NLO            &0.165542      & 0.165538         \\
NNLO           &0.164938      &0.164934         \\
NNNLO          &0.164671      & 0.164667        \\
\hline\hline
\end{array}
\]
\caption{\label{Tab4}The quantity 
$I^{(1)}_{\tau b}$ from eq.~(\ref{fortable4})
in different orders of strong $\beta$-function
and with or without the EM contribution to 
the strong $\beta$-function.}
\end{table}

\noindent For the scale $M_Z$ the error due to running is
estimated by 
\[
\delta\Delta^{hadcor}_{\tau Z}|_{a_s}
=\frac{11}{3}\delta I_{\tau Z}^{(1)}(a_s(m_b))
=13.6\delta a_s=0.074
\]
which leads to 
\[
\delta\Delta^{(5)}(M_Z^2)
=\pm 0.078_{light}\pm 0.330_c  \pm 0.030_{b} \pm 0.074_{run} 
\]
\be
\label{errFinal}
= \pm 0.152_{light+run} \pm 0.330_c  \pm 0.030_{b}\, .
\ee
Adding independent errors in quadrature one has
\[
\delta\Delta^{(5)}(M_Z^2)=\pm 0.3646
\]
and 
\[
0.3646/(3\pi)=0.03869
\]
These estimates give the error for the coupling at $M_Z$
\be
\label{finalZres}
\frac{1}{\bal^{(5)}(M_Z^2)}=127.843\pm 0.039\, .
\ee
Eq.~(\ref{finalZres}) is a final result.
However, it cannot be directly compared with the results of 
the standard analyses because the quantity in eq.~(\ref{finalZres}) 
is defined in a different scheme.
We consider the uncertainty in eq.~(\ref{finalZres})
(the part of which related to the running 
is estimated analytically for transparency) 
as rather conservative.

\section{Comparison with other schemes}
With the number from eq.~(\ref{finalZres})
one can find the on-shell parameter
$\al_{os}$ at $M_Z$ used in the literature. 
Indeed, because of the relation
\begin{equation}
  \label{match11}
\frac{3\pi}{\al_{os}({\bf q^2})}
=\frac{3\pi}{\al}+\Pi_{os}({\bf q^2})
=\frac{3\pi}{\bal^{(5)}(\mu^2)}+\Pi^{(5)}(\mu^2,{\bf q^2})\, .
\end{equation}
one has to compute $\Pi^{(5)}(\mu^2,{\bf q^2})$
in $n_f=5$ effective theory at the point $q^2\sim M_Z^2$.
Note that we have restored a notation $q^2$ in addition to 
the usual quantity ${\bf q^2}$: the new variable
$q^2$ will be used in Minkowskian domain.
For computing the leading part of 
$\Pi^{(5)}(\mu^2,{\bf q^2})$ in the kinematical range 
$\mu^2\sim {\bf q^2}\sim M_Z^2$ one can consider 
all five active quarks ($u$, $d$, $s$, $c$, $b$) and all three
leptons as massless and use 
eq.~(\ref{Pilight}) with the only change because of a different 
number of active quarks which is now 5 instead of 3.
This change affects only ${\cal O}(a_s^2)$ order in 
eq.~(\ref{Pilight}) and 
changes nothing for leptons in the NLO approximation.
One has for the generic light polarization function 
of quarks 
\[
\Pi^{light-{n_f}}(\mu^2,{\bf q}^2)
=\ln \frac{\mu^2}{{\bf q}^2}+\frac{5}{3}
+a_s\left(\ln \frac{\mu^2}{{\bf q}^2}+\frac{55}{12}-4\zeta(3)\right)
\]
\[
+a_s^2\left\{\frac{\beta_0(n_f)}{2}\ln^2 \frac{\mu^2}{{\bf q}^2}
+\left(\frac{365}{24}-\frac{11}{12}n_f
-4\beta_0(n_f)\zeta(3)\right)\ln \frac{\mu^2}{{\bf q}^2}
\right.
\]
\be
\label{PilightNf}
\left.
+\frac{41927}{864}-\frac{3701}{1296}n_f
-\left(\frac{829}{18}-\frac{19}{9}n_f\right)\zeta(3)+\frac{25}{3}\zeta(5)
\right\}
\ee
with $\beta_0(n_f)=(11-2/3 n_f)/4$. 
For a more accurate evaluation of $\Pi^{(5)}(\mu^2,{\bf q^2})$
at the scale $M_Z$ 
we retain the leading corrections due to 
the $b$-, $c$-quark and $\tau$-lepton masses
and the leading correction due to the top quark contribution.
One finds
\[
\Pi^{(5)}(\mu^2,{\bf q}^2)=3\Pi^{light-lept}(\mu^2,{\bf q}^2)+
\frac{11}{3}\Pi^{light-quark}(\mu^2,{\bf q}^2)
\]
\be
\label{mZexpPol}
-\frac{1}{3}\frac{6 m_b^2}{{\bf q}^2}-\frac{4}{3}\frac{6 m_c^2}{{\bf q}^2}
-\frac{6 M_\tau^2}{{\bf q}^2}
+\Delta_{(t)}\Pi^{(5)}(\mu^2,{\bf q}^2)\, .
\ee
Note that the power correction due to a quark mass
is exact up to ${\cal O}(a_s^2)$ order when expressed
through the pole mass.  
The corrections due to the top quark contribution
for the quantity $\Pi^{(5)}(\mu^2,q^2)$ at ${\bf q^2}\approx M_Z^2$
can be computed as a power series in ${\bf q^2}/m_t^2$;
the expansion parameter ${\bf q^2}/m_t^2$ is small
at the point ${\bf q^2}=M_Z^2$ for $m_t=175~{\rm GeV}$. 
Indeed, retaining only the leading term and first corrections one has
\be
\label{topcorr} 
\Delta_{(t)}\Pi^{(5)}(M_Z^2,{\bf q}^2)
=-\frac{4}{15}\frac{{\bf q}^2}{m_t^2}
\left\{1+\frac{410}{81} a_s^{(5)}(M_Z^2)
-\frac{3}{28}\frac{{\bf q}^2}{m_t^2}
\right\}\, .
\ee
A typical expansion in eq.~(\ref{topcorr}) reads
\be
\label{typtop}
\Delta_{(t)}\Pi^{(5)}(M_Z^2,M_Z^2)
=-0.0724 - 0.0138_{a_s}  + 0.0021_{M_Z}=-0.0841
\ee
with obvious notation indicating the origin of different contributions.
We do not take into account bosons therefore the $W$-boson loops should be
analyzed separately.

To calculate the on-shell coupling $\al_{os}({\bf q}^2)$
at the scale $M_Z$ using eq.~(\ref{match11})
one can use either ${\bf q}^2=M_Z^2$
(Euclidean definition) 
\cite{Korner:1999ke} or $q^2=-{\bf q}^2=M_Z^2$ with taking the real part
of the correlator (Minkowskian definition).
The Minkowskian definition is usually discussed in the literature.
Note that we calculate not the $e^+e^-$ scattering amplitude at 
the total energy $M_Z$ ($q^2=M_Z^2$) which
definitely should be taken at a physical point on the cut in the case
of cross section calculations, but 
the coupling constant which parameterizes this amplitude at the scale
$M_Z$. Within the RG approach the scale of the parameters of an 
effective theory valid in a given energy range should not coincide
with any actual physical value of the energy or momentum squared
(see e.g. \cite{krajw12}). 

First we use a Euclidean definition for the on-shell coupling
which is consistently perturbative 
and requires computation of the correlator 
$\Pi^{(5)}(M_Z^2,{\bf q}^2)$
in a deep Euclidean domain for ${\bf q}^2=M_Z^2$.
Using eqs.~(\ref{mZexpPol},\ref{topcorr}) one finds an expansion 
\[
\Pi^{(5)}(M_Z^2,M_Z^2)=11.1111 - 0.03097_{a_s}
 + 0.00112_{a_s^2}
\]
\[
- 0.00168_{EM}-0.00554_b- 0.00304_c- 0.00228_\tau-0.0841_t
\]
\be
\label{pi5exp}
=11.0796-0.01086_{bc\tau}-0.0841_t\, .
\ee
Note that the EM correction is numerically of the order $a_s^2$. 
Still these corrections are very small. 
From eq.~(\ref{match11}) we find
\[
\frac{3\pi}{\al_{os}(M_Z^2)}
=\frac{3\pi}{\bal^{(5)}(M_Z^2)}+\Pi^{(5)}(M_Z^2,M_Z^2)
\]
\[
=\frac{3\pi}{\al}-86.6384+11.0796-0.01086_{bc\tau}-0.0841_{t}
\]
\[
=-75.5588-0.01086_{bc\tau}-0.0841_t\, .
\]
For clarity we retain the contribution of power corrections separately
for further comparison with the results in Minkowskian domain.
One has numerically 
\[
\frac{1}{\al_{os}(M_Z^2)}
=\frac{1}{\al}-(75.5588+0.01086_{bc\tau}+0.0841_t)/(3\pi)
\]
\[
=137.036-8.0271=129.009\, .
\]
Because the error estimate in eq.~(\ref{finalZres}) 
is not affected by this change of scheme (too small and rather
precise contributions are added),
the final result for the on-shell coupling within 
Euclidean definition reads 
\be
\label{finalEucl}
\frac{1}{\al_{os}(M_Z^2)}=129.009\pm 0.039\, .
\ee
However, the reference values for the on-shell coupling
available in the literature are
given in Minkowskian domain for $q^2=-{\bf q}^2=M_Z^2$, i.e.
for the real part of the correlator 
$\Pi^{(5)}(M_Z^2,q^2)$ computed on the physical cut.
Within the approximation used, going to the Minkowskian domain of 
momenta $q^2$ changes only the $a_s^2$ order term and power corrections 
in eq.~(\ref{pi5exp}). Indeed, in eq.~(\ref{PilightNf}) the only term
which is numerically affected by the transition to the Minkowskian domain is
$\ln^2(\mu^2/{\bf q}^2)$
with the following change as compared to the Euclidean result
\be
{\rm Re}~\left\{\ln^2\left(\frac{\mu^2}{-M_Z^2}\right)\right\}
=\ln^2\left(\frac{\mu^2}{M_Z^2}\right)-\pi^2 \, .
\ee
Instead of eq.~(\ref{pi5exp}) one finds
\[
{\rm Re}~\left\{\Pi^{(5)}(M_Z^2,-M_Z^2)\right\}
=11.0796
\]
\be
\label{topMinkOur}
-0.04893_{\pi^2}+0.01086_{bc\tau Mink}+0.08828_{tMink}
=11.1298
\ee
and 
\[
{\rm Re}~\frac{1}{\al_{os}(-M_Z^2)}
=137.036-(86.6384-11.1298)/(3\pi)=129.024\, .
\]
The final result for the on-shell coupling with 
Minkowskian definition is 
\be
\label{finalMink}
{\rm Re}~\frac{1}{\al_{os}(-M_Z^2)}=129.024\pm 0.039\, .
\ee
The difference between the central values for the couplings 
in eq.~(\ref{finalEucl})
which gives the Euclidean definition and in eq.~(\ref{finalMink})
which gives the Minkowskian definition is 0.015.
Note that the Euclidean definition was considered 
in ref.~\cite{Korner:1999ke} where the numerical difference 
about $0.02$ from the Minkowskian definition
has been found from rather a simplified treatment.
It is close to the present more accurate result 0.015.
Note that the point $M_Z$ is safe for the PT calculation in 
Minkowskian
domain for the approximation used (no singularities of the spectrum
near this point).
At other points it is not so even in the approximation we work. 
For instance, if the $\Upsilon$-resonance
mass $m_\Upsilon$ is taken as a reference scale then 
the Euclidean definition is equally applicable at this point 
being still perturbative while 
the Minkowskian definition faces the problems
that the polarization function on the cut is not smooth. 
A phenomenological approach based on direct integration of data 
fails because of the fast
change of the spectrum at the location of the $\Upsilon$
resonance which makes the integration
with principal value prescription for regularizing the singularities
ill-defined. A theory based approach within PT 
fails at the point $m_\Upsilon$ because PT calculations for the correlator 
near the threshold on the physical cut
($m_\Upsilon\sim 2m_b$) are not reliable. Therefore,
the Minkowskian definition is not uniformly applicable at 
every scale.

The present paper result given in eq.~(\ref{finalMink}) 
differs from some recent determinations
based on the use of 
experimental data for integration over the low energy region.
For the result of ref.~\cite{Schilcher} 
\be
\label{SchilcherRes}
{\rm Re}~\frac{1}{\al_{os}(-M_Z^2)}=128.925\pm 0.056
\ee
the number obtained in the present paper and given in 
eq.~(\ref{finalMink})
almost touches the reference value  eq.~(\ref{SchilcherRes})
within $1\sigma$ ($\sigma$ is a standard deviation). 
The results of some other groups are concentrated around 
the same central value as in 
eq.~(\ref{SchilcherRes}) but with essentially smaller errors.
For further comparison, we use the result of ref.~\cite{Kuhn}
\be
\label{alKu}
{\rm Re}~\frac{1}{\al_{os}(-M_Z^2)}=128.927\pm 0.023 \, .
\ee
The difference between the value from eq.~(\ref{finalMink}) and 
the reference result in eq.~(\ref{alKu}) is 
$129.024-128.927=0.097$ which constitutes 2-4$\sigma$
and can be significant.
Therefore we discuss the difference in more detail.

The usual parameterization of the fermionic contributions to the 
on-shell running EM coupling at $M_Z$ reads
\[
{\rm Re}~\frac{1}{\al_{os}(-M_Z^2)}=
\frac{1}{\al}\left(1-\Delta{\al}_{\rm lep}-\Delta{\al}^{(5)}_{\rm had}
-\Delta{\al}_{\rm top}\right)
\]
(note that 
\[
{\rm Re}~\left(\frac{1}{\al_{os}(-M_Z^2)}\right)
\ne \frac{1}{{\rm Re}~\al_{os}(-M_Z^2)}
\]
though the difference is tiny). 
The total leptonic contribution to the electromagnetic coupling
constant 
at $M_Z$ given in the last column of Table \ref{Tab1}
reads
\[
\Delta^{lep}(M_Z^2)=45.681.
\]
The leptonic part of $\Pi^{(5)}(M_Z^2,-M_Z^2)$
reads
\[
{\rm Re}~\left\{\Pi^{(5)lep}(M_Z^2,-M_Z^2)\right\}
=3{\rm Re}~\left(\Pi^{light-lept}(M_Z^2,-M_Z^2)\right)
+\frac{6 M_\tau^2}{M_Z^2}
\]
\[
=4.9988 + 0.0023=5.0011\, .
\]
The leading order contribution is equal to 5 while 
the EM and $\tau$-lepton mass corrections are small.
For the total leptonic contribution to the on-shell coupling in
the Minkowskian domain one finds
\[
\Delta{\al}_{\rm lep}=\frac{\al}{3\pi}(45.681-5.001)=314.974\times 10^{-4}
\]
which is close to the number of ref.~\cite{Kuhn}
\[
\Delta{\al}_{\rm lep}|_{\rm ref}=(314.19+0.78)\times 10^{-4}
=314.97\times 10^{-4}\, .
\]
For the top contribution in the Minkowskian domain 
one finds from
eq.~(\ref{topcorr}) (see also eq.~(\ref{typtop}))
\[
\Delta{\al}_{\rm top}=\frac{\al}{3\pi}(-0.0883)= -0.68\times 10^{-4}
\]
while the number of ref.~\cite{Kuhn} with a more accurate account for
the higher order corrections is
\[
\Delta{\al}_{\rm top}|_{\rm ref}=-0.70\times 10^{-4}\, .
\]
The difference is small and is neglected.
From the numerical value given in eq.~(\ref{finalMink})
the total contribution of fermions into the shift of the EM coupling 
is determined to be 
\[
\Delta{\al}_{\rm lep}+\Delta{\al}_{\rm top}+\Delta{\al}^{(5)}_{\rm had}
=1-\al\cdot (129.024\pm 0.039)=0.0584664\pm 0.000285\, .
\]
Taking $\Delta{\al}_{\rm lep}$ and $\Delta{\al}_{\rm top}$ as exact
quantities (no errors)  
one finds the following numerical value for 
$\Delta{\al}^{(5)}_{\rm had}$ 
\bea
\label{d5I}
\Delta{\al}^{(5)}_{\rm had}
&=&(0.0584664\pm 0.000285)-0.031497+0.000068\nn \\
&=&(270.37\pm 2.85)\times 10^{-4}
\eea
while the result of ref.~\cite{Kuhn} is 
\be
\label{d5K}
\Delta{\al}^{(5)}_{\rm had}|_1=(277.45\pm 1.68)\times 10^{-4}\, ,
\ee
and the number of ref.~\cite{Schilcher} is 
\be
\label{d5S}
\Delta{\al}^{(5)}_{\rm had}|_2=(277.6\pm 4.1)\times 10^{-4}\, .
\ee
The difference between the central values given 
in eq.~(\ref{d5I}) and eqs.~(\ref{d5K},\ref{d5S}) 
is about 2-4$\sigma$ depending on the numerical value of the error
quoted
\[
277.45-270.37 
= 
7.08 \approx 2.5\cdot 2.85  \approx 4.3\cdot 1.68 \approx 1.7\cdot 4.1\, .
\] 
This difference can be significant.
Therefore, we discuss the sensitivity of our prediction 
(\ref{d5I}) (and of eq.~(\ref{finalZres}) from which 
it is uniquely obtained) to the numerical values of parameters used
in the theoretical calculation of the present paper.
If $m_s=0$ then the $\rho$- and $\varphi$-channels should be theoretically 
degenerate because there is no reason for the difference.
This means that besides vanishing explicit corrections due to $m_s^2$ 
in eq.~(\ref{strSpol})
one should identify $m_{Rs}$ with the resonance in the nonstrange 
channel, i.e. numerically substitute $m_{Rs}=m_\rho$
into the solution for the IR modifying parameters in eqs.~(\ref{sFESR}).
With such changes one finds 
the result for $\Delta^{uds}(M_\tau^2)$ in the form 
\be
\label{ms0ch}
\Delta^{uds}(M_\tau^2)|_{m_s=0} = 10.23
\ee
that generates a numerical shift about $0.3$
in the value of $\Delta^{uds}(M_\tau^2)$
as compared to the result for nonvanishing strange quark mass
in eq.~(\ref{finallight}). Note that if the direct integration of low
energy data is used then the full dependence of the results on 
$m_s$ is lost. Only the PT high energy tail depends explicitly on $m_s$ 
but this dependence is weak.
The change in $\Delta{\al}^{(5)}_{\rm had}$ corresponding to 
the result in eq.~(\ref{ms0ch}) is
\[
\Delta{\al}^{(5)}_{\rm had}|_{m_s=0}-\Delta{\al}^{(5)}_{\rm had}
=0.3\frac{\al}{3\pi} =0.000232=2.3\times 10^{-4} \, .
\]
The use of the numerical value $m_c=1.6~{\rm GeV}$ 
for the $c$-quark mass instead of $m_c=1.777~{\rm GeV}$
generates the $0.33$ shift in the value of quantity
$\Delta^{c}(M_\tau^2)$
that leads to the following change of $\Delta{\al}^{(5)}_{\rm had}$
\[
\Delta{\al}^{(5)}_{\rm had}|_{m_c=1.6}- \Delta{\al}^{(5)}_{\rm had}
=0.33\frac{\al}{3\pi}=0.000256=2.6\times 10^{-4}\, .
\]
Note that this change cannot be found if the direct integration of 
contributions of actual charmonium resonances is performed.
The total shift due to such a change in these parameters 
compared to eq.~(\ref{d5I}) is 
\bea
\label{shmcms}
\Delta{\al}^{(5)}_{\rm had}|_{m_s=0,m_c=1.6}
&=&(270.37+2.3+2.6)\times 10^{-4}\nn \\
&=&(275.27\pm 2.85)\times 10^{-4}\, .
\eea
The change of the numerical value of the strong coupling constant 
$\al_s$ from $\al_s(M_\tau^2)=0.318$ to 
$\al_s(M_\tau^2)=0.335$ gives a 0.152 shift in 
$\Delta^{(5)}(M_Z^2)$ (according to our error
estimates in eq.~(\ref{errFinal})) to end up with 
\bea
\label{shmcmsas}
\Delta{\al}^{(5)}_{\rm had}|_{m_s=0,m_c=1.6,\al_s=0.335}
&=&(270.37+2.3+2.6+1.2)\times 10^{-4}\nn \\
&=&(276.47\pm 2.85)\times 10^{-4}\, .
\eea
This result agrees with other estimates within $1\sigma$.
The set of numerical values for the relevant parameters used in 
eq.~(\ref{shmcmsas}) is rather close to the set used 
for obtaining the value in eq.~(\ref{d5I})
($m_s=130~{\rm MeV}$, $m_c=1.777~{\rm GeV}$,
$\al_s^{(3)}(M_\tau^2)=0.318$).
The total shift in $\Delta{\al}^{(5)}_{\rm had}$
for the new set of parameters 
in eq.~(\ref{shmcmsas}) is larger than the total error 
given in eq.~(\ref{d5I}) because the total error is computed
in quadrature and the change of the spectrum due to $m_s=0$
(which makes all three light quark channels degenerate)
has not been included into the total error.
To definitely distinguish between the results of 
eq.~(\ref{shmcmsas}) and eq.~(\ref{d5I}) more precise 
numerical values of parameters are necessary. 

Within the present paper approach we use virtually no real data on
cross sections but rely on the numerical values of several theoretical
parameters which are extracted from such data.
These parameters are the strong coupling constant, quark masses,
vacuum condensates. It is generally believed that
the real data can be properly described with these parameters 
if theoretical formulas are
sufficiently accurate. In the case of computing 
the hadronic contribution into the photon vacuum
polarization function
in Euclidean domain a theoretical description 
is pretty accurate because PT is applicable
and very precise -- in fact, the PT results in this area 
are almost the best ones available among all PT calculations.
An additional reason for such a high theoretical 
precision is that for the calculation of 
$\Delta{\al}^{(5)}_{\rm had}$ in Euclidean domain 
one extracts only very
general information encoded in the data -- just the integral over the
entire spectrum with a smooth weight function and no details of the
behavior over specific energy regions. This is
the situation when global
duality, which is exact by definition (hadron and quark descriptions 
are supposed to be exactly equivalent in principle), is applicable
and is under a strict control numerically within PT.
However, our calculation shows that at the present level of precision
the result for $\Delta{\al}^{(5)}_{\rm had}$ 
is rather sensitive to the numerical values of the parameters
$m_c$ and $a_s$ which should be fixed from the data. 
The uncertainties of these parameters 
can be reduced both with better data
and better theoretical formulas for extracting
the numerical values for these parameters 
from the data (especially $m_c$) while the
theoretical framework for the calculation of the hadronic
contribution itself is already very precise.
Using the result given in eq.~(\ref{d5I}) and formulas for radiative
corrections to the Weinberg angle from ref.~\cite{hihi}
(assuming Minkowskian definition for the on-shell coupling)
we find that the central value of the Higgs-boson mass moves from 
the reference value $M_H=100~{\rm GeV}$ for 
$\Delta{\al}^{(5)}_{\rm had}=280.0\times 10^{-4}$ to 
$M_H=191~{\rm GeV}$ for
the value $\Delta{\al}^{(5)}_{\rm had}=270.37\times 10^{-4}$ found in the
present paper.

\section{Conclusion}
The technique of calculating $\Delta{\al}^{(5)}_{\rm had}$
within dimensional regularization and minimal subtraction
is straightforward in PT.
It heavily uses the renormalization group
which is the most powerful
tool of modern high precision analyses in particle phenomenology.
Because PT is not applicable only at 
low energies one should modify only the IR region of integration
for light quarks:
a numerical integration of data at energies higher 
than $2\div 3~{\rm GeV}$
is equivalent to the theoretical calculation in PT
if both data and theory are properly treated.
The present calculation uses virtually no explicit scattering 
data but the values of the lowest
resonance masses for the light-quark vector channels.
Other experimental information is encoded through the numerical values
of the coupling constant, quark masses, some vacuum
condensates.
 
Minkowskian definition of the on-shell coupling constant
is deficient and not applicable at some points. 
Both $\MSsch$ and on-shell coupling constants in the Euclidean domain 
can be reliable determined 
with the use of theoretical formulas already established
in high orders of PT. 
In view of future high precision tests of SM at $M_Z$ and 
two-loop calculations for the observables in this region it seems
that the parameterization of the theory 
with the running EM coupling in the $\MSsch$ scheme is most promising.

The main uncertainty of the hadronic contribution
into the running EM coupling constant at $M_Z$  
comes from the error of the numerical value of the 
$c$-quark mass $m_c$.
The uncertainty of $a_s$ is less important.
Unfortunately, the $c$-quark mass is a quantity which 
is very complicated to study.
The reason for that is its numerical value close to 
the strong interaction scale of
the order of $\rho$-meson or proton mass. Therefore, $m_c$
should be treated exactly in theoretical formulas, 
almost no simplifying approximation is applicable in
the kinematical range of energies of order $m_c$. 
The presence of a mass usually makes
the loop calculations within PT technically more difficult. 
Near the $c\bar c$ production 
threshold where the mass is essential and 
where its numerical value can be reliably extracted
from accurate experimental data, the Coulomb interaction 
is enhanced that 
requires to take it into account exactly 
while the $c$-quark mass is too
small for NRQCD to work well.
Finally, the nonPT corrections due to vacuum condensates 
within OPE are essential 
numerically in this energy range
but they are not well known because they are
given by the gluonic operators \cite{charm}. And though 
the coefficient functions of the relevant operators up to 
dimension eight are calculated \cite{radcharm},  
the numerical values of their vacuum condensates 
are poorly known.
These reasons make the accurate determination of the $c$-quark mass difficult.
The uncertainty related to the contribution of the $c$ quark 
to the hadronic vacuum polarization
is additionally enhanced because of the $c$-quark large electric charge.
For the $b$ quark, for instance,
all the above problems are much less severe.
Therefore, the $c$-quark physics plays an essential part  
in the Higgs search through radiative corrections.

\section*{Acknowledgement}
I am indebted to K.G.~Chetyrkin for discussion
of various aspects of many-loop calculations and correspondence.
Discussions with F.~Jegerlehner, J.G.~K\"orner, J.H. K\"uhn,
and K.~Schilcher are thankfully acknowledged.
The work is partially supported 
by the Russian Fund for Basic Research under contract
99-01-00091 and by Volkswagen 
Foundation under contract No.~I/73611. 
Presently A.A.~Pivovarov is an Alexander von Humboldt fellow.

\end{document}